%% file: usenix_main.tex
\documentclass[letterpaper,twocolumn,10pt]{article}
\usepackage{usenix2019_v3}

\usepackage{amsmath}
\usepackage{amsfonts}
\usepackage{natbib}
\usepackage{amsmath}
\usepackage{amsthm}
\usepackage[frozencache,cachedir=minted-cache]{minted}

\usepackage{graphicx}
\usepackage{booktabs}
\usepackage{fancyvrb}
\usepackage{nicefrac}
\usepackage{tcolorbox}
\usepackage{xcolor}
\usepackage{multirow}
\usepackage{adjustbox}
\usepackage{algorithm}
\usepackage{algpseudocode}
\usepackage{tcolorbox}
\usepackage{cleveref}

\theoremstyle{definition}

\usepackage{caption}

\newcommand{\myexample}[2]{
    \begin{tcolorbox}[colback=black!5!white,colframe=black,title={#1}]
        #2
    \end{tcolorbox}
}

\usepackage{tikz}
\newcommand*\circled[1]{\tikz[baseline=(char.base)]{
            \node[shape=circle,fill,inner sep=2pt] (char) {\textcolor{white}{#1}};}}

\newcommand{\eg}{\textit{e.g\@.}}

\newcommand{\ie}{\textit{i.e\@.}}
\newcommand{\ab}{\textit{architectural backdoor\@}}
\newcommand{\abs}{\textit{architectural backdoors\@}}
\newcommand{\Ab}{\textit{Architectural backdoor\@}}
\newcommand{\Abs}{\textit{Architectural backdoors\@}}

\newcommand{\var}{\texttt}
\newcommand{\func}{\textrm}

\begin{document}
%-------------------------------------------------------------------------------

%don't want date printed
\date{}

% make title bold and 14 pt font (Latex default is non-bold, 16 pt)
\title{\Large \bf Architectural Neural Backdoors from First Principles}

\author{}
%for single author (just remove % characters)
\author{
{\rm Harry Langford}\\
University of Cambridge
\and
{\rm Ilia Shumailov}\\
University of Oxford
\and
{\rm Yiren Zhao}\\
Imperial College London
\and
{\rm Robert Mullins}\\
University of Cambridge
\and
{\rm Nicolas Papernot}\\
University of Toronto
} % end author

\maketitle

\input{sections/abstract}
\input{sections/introduction}
\input{sections/related}
\input{sections/methodology}

\input{sections/evaluation}

\input{sections/discussion}

\input{sections/conclusion}
%-------------------------------------------------------------------------------

\input{sections/acks}
\bibliographystyle{abbrvnat}
\bibliography{bibliography}

\input{sections/appendix}

%%%%%%%%%%%%%%%%%%%%%%%%%%%%%%%%%%%%%%%%%%%%%%%%%%%%%%%%%%%%%%%%%%%%%%%%%%%%%%%%
\end{document}

%% file: sections/abstract.tex
\begin{abstract}
While previous research backdoored neural networks by changing their parameters, recent work uncovered a more insidious threat: backdoors embedded within the definition of the network's architecture. This involves injecting common architectural  components, such as activation functions and pooling layers, to subtly introduce a backdoor behavior that persists even after (full re-)training. However, the full scope and implications of architectural backdoors have remained largely unexplored. \citet{bober2023architectural} introduced the first architectural backdoor; they showed how to create a backdoor for a checkerboard pattern, but never explained how to target an arbitrary trigger pattern of choice. In this work we construct an arbitrary trigger detector which can be used to backdoor an architecture with no human supervision. This leads us to revisit the concept of~\abs~and taxonomise them, describing 12 distinct types. To gauge the difficulty of detecting such backdoors, we conducted a user study, revealing that ML developers can only identify suspicious components in common model definitions as backdoors in 37\% of cases, while they surprisingly preferred backdoored models in 33\% of cases. To contextualize these results, we find that language models outperform humans at the detection of backdoors. Finally, we discuss defenses against architectural backdoors, emphasizing the need for robust and comprehensive strategies to safeguard the integrity of ML systems.
\end{abstract}

%% file: sections/introduction.tex
\input{./sections/diagrams/paths.tex}

\section{Introduction}
Backdoors are one of the main risks to machine learning (ML) security. Early work demonstrated that by changing training data an adversary can introduce backdoors into models, forcing them to learn attacker-controlled patterns called `triggers'~\citep{gu2019badnets}. Training with such triggered data allows the attacker to then control model inference by simply adding the trigger to otherwise benign data. At the same time, a parallel research stream looked into backdooring ML models without relying on changing underlying data. At the time of writing, the community is aware that backdoors can hide in \eg~quantisation~\citep{ma2023quantization}, loss functions~\citep{bagdasaryan2021blind}, ML compilers~\citep{clifford2022impnet}, data samplers~\citep{shumailov2021manipulating}, and augmentation routines~\citep{rance2022augmentation}.
More recently it was demonstrated that model architectures are also vulnerable to architectural backdoors~\citep{bober2023architectural}. In this setting, an attacker changes the model definition only using common ML components to introduce explicit bias into the architecture such that training or fine-tuning of the architecture results in a backdoored model. 

\citet{bober2023architectural} showed a proof-of-concept Model Architectural Backdoor (MAB), thereby demonstrating that such backdoors are \textbf{\textit{possible}}. MAB described a detector that could be used to probabalistically find a checkerboard pattern in the input. Importantly, MAB was limited in that it activated only some of the time, had a chance to be disabled during training, and most importantly, did not describe a way to construct arbitrary trigger pattern detectors. In this paper we address the shortcomings of the original design and demonstrate that the original MAB design is an example of a backdoor from a larger backdoor family with different trade-offs. 

This work is structured as follows. First, we show how one can construct a trigger detector for an arbitrary pattern of choice~\ie~we show that arbitrary patterns can be used as triggers with provable guarantees that they remain in the network, even after training from scratch. In other words, we can add commonly used ML components into neural networks to make them behave deterministically in the presence of a given trigger, all without any learned parameters -- this means that even if models are retrained from scratch they will still be provably backdoored. We do this by assembling the trigger detector from the basic logic blocks approximated using basic ML operators~\eg~activations. Next, we show that backdoors are easy to inject into neural networks both before and after training, with no impact on performance of the final network in either case. We then taxonomise~\abs~and show that they outperform prior literature in both accuracy and survivability, since our backdoors are injected as deterministic boolean functions that always evaluate in the same way. We run a user study to investigate if \abs~can be detected by everyday ML practitioners. We find that they successfully identify backdoors in common model definitions as suspicious only in 37\% of the time, while backdoored models are blindly preferred by the developers in 33\% of pairwise comparisons. Finally, we outline how to check if a given architecture may be vulnerable and disable the attack. Overall, we make the following contributions:
\begin{itemize}
    \item We describe a family of architectural backdoors, taxonomising them along three main dimensions: where they detect the trigger, how they propagate the trigger signal, and how they integrate the signal back into the network. We show that architectural backdoors can target arbitrary triggers and have provable operational guarantees.
    \item We run a user study and find that human preference for model architecture is influenced more by coding style rather than the presence of backdoors. More precisely, backdoored models are preferred 33\% of the time, while backdoors are detected only 37\% of the time. 
    \item We explore methods and tools that can be employed to effectively defend against~\abs, covering visual inspections, provenance, architectural sandboxing, and access controls.
\end{itemize}

%% file: sections/diagrams/paths.tex
\begin{figure*}
    \centering
    \includegraphics[width=410pt]{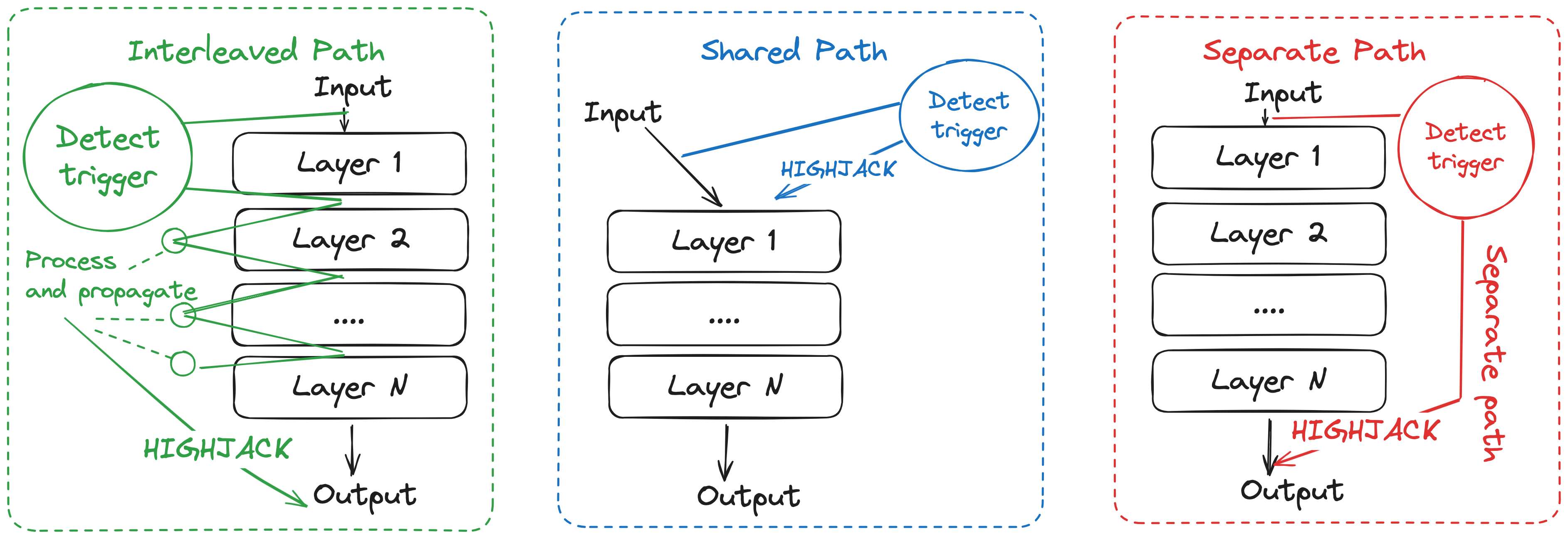}
    \caption{A visualisation of the differences between the mode of transmission of the trigger signal from input to output. Left shows interleaved path merging and being separated from the main network. Centre shows shared path directly augmenting a representation of the data. Right shows separate path reading and augmenting a later part of the network.}
    \label{fig:overall_flow}
\end{figure*}

%% file: sections/related.tex
\section{Related Work}

\noindent We first describe the literature on classic software backdoors, their effects on software supply chains, and the defenses against such attacks. Then, we cover ML specific backdoors.

\noindent \textbf{Software Backdoors} Backdoor attacks become possible when an attacker gains control of a part of the overall supply chain. Growing complexity of modern software supply chains makes the potential security risks associated with backdoor attacks ever more realistic. Attacks involving backdoors exist across the whole software ecosystem, ranging from peripheral devices to operating systems~\citep{sshbackdoor,eventstream,samsung}. A major mitigation strategy for backdoor attacks involves analysing the dependency chains of the software in use. However, this task is particularly challenging with modern software, which entails not only a wide (numerous imported modules), but also a deep (multiple levels of dependencies) dependency graph~\citep{boucher2023automatic,balliu2023challenges,xia2023empirical}. 

\noindent \textbf{ML Backdoors} ML is bringing even more complexity into our supply chains: pipelines use ML components, which are built using insecure code with proprietary data, are fully or partially generated by ML; and the generated code is trained on by other ML models exacerbating problems further~\citep{shumailov2023curse}. For example, PyTorch-nightly discovered a compromised dependency chain, leading many users having to reinstall the package in late 2022 \citep{torchbreak}. In another instance, Huggingface's load by name feature was used to perform `AIJacking' attacks, where names of previously registered models, were re-registered by malicious actors~\citep{torchbreak4legit}. In some cases custom arbitrary code can be injected into model checkpoints~\citep{lobotoml}.  Finally, recently supply chain attacks were reported against the \texttt{PyTorch} package by abusing GitHubs runners feature and allowed for arbitrary code injection~\citep{torchbreak3young,torchbreak2stawinski}. 

The first neural network-specific backdoor attacks were demonstrated by \citet{gu2019badnets}, who revealed that attackers, who have the ability to modify the training data, can force deep neural networks to learn specific attacker-controlled features. Since then, considerable advancements have been made in these attacks.
\citet{saha2020hidden} developed a strategy for trigger design, wherein the attacker conceals the trigger within the poisoned data and keeps it hidden until the time of testing. \citet{shafahi2018poison} extended the poisoning attacks by utilising exclusively data with clean labels. Additionally, \citet{salem2022dynamic} enhanced the efficiency of triggers. 

The aforementioned work focuses on data-based approaches, striving to enhance the stealthiness and effectiveness of triggers assuming the attacker's ability to manipulate training data. However, many other studies have explored different attack vectors within the machine learning supply chain. \citet{hong2022handcrafted} devised an approach where they manually crafted weights to create a more potent backdoor effect. \citet{ma2023quantization} showcased backdoors can be injected through the weight quantisation process. \citet{bagdasaryan2021blind} injected a backdoor through a loss function. \citet{shumailov2021manipulating} on the other hand, introduced a technique to backdoor models by infecting only the training data sampler and arranging the data in a specific order during training to simulate the effect of creating triggers. \citet{rance2022augmentation} exploited the data augmentation process to insert backdoors into the models they targeted.
\citet{clifford2022impnet} demonstrated that backdoors can also exist in ML compilers, and the detection of such backdoors is almost impossible. \citet{warnecke2023evil} showed that hardware can also be used to launch backdoor attacks. 

Our work aligns closely with a specific threat model in which the attacker introduces \textit{architectural backdoors} into a network provided to a user. The user then trains the network from scratch using their own data and subsequently deploys the trained model. 
The concept of an~\ab~was introduced by~\citet{bober2023architectural}. In our paper, we generalise and taxonomise this attack, and describe it as a family of diverse architectural backdoors. Furthermore, we evaluate and verify their effectiveness through an in-person user study involving participants with diverse levels of expertise in ML.
\citet{pang2023the} since showed that such backdoors can be discovered with NAS. 

%% file: sections/methodology.tex
\section{Methodology}

In this section we first describe the previous design of~\abs~and in what settings they are expected to work. We then turn to describing how to generalise the initial design to build an arbitrary trigger detector. Further, we  taxonomise them along three main dimensions.

\subsection{Primer on backdoors} 

\textbf{Definition 1} A neural network \textit{\textbf{backdoor}} is a hidden unauthorised functionality in a neural network that can be activated by the attacker using a special trigger providing them with some control over the model. A \textit{\textbf{trigger}}, in turn, is a modification to the original input that is used to activate a backdoor. 

\noindent
\textbf{Definition 2} 
\textbf{\textit{Architectural backdoor}} is a type of a machine learning backdoor that exists in an architecture definition of a neural network. The backdoor then \textit{implicitly or explicitly} \textit{biases} the architecture to exhibit adversary-chosen behaviours in presence of adversary-chosen triggers.

We call our backdoor an \ab~-- these are backdoors which are embedded into the architecture of a model. They are made by augmenting the architecture of an existing network to provide an adversary with some notion of control over the model. In this work, we focus on injecting deterministic class backdoors into the model, but in practice other goals may be pursued~\eg~biasing the models towards a certain class, stopping the model from training, activating a hardware backdoor, or more generally denying service. Since the backdoors are embedded into the architecture, they can be weight-invariant and dataset-invariant meaning that they can work on all models on all datasets, regardless of training. 

Specifically, we assume a benign classifier $f: \mathbb{R}^n \rightarrow \mathbb{R}^m$ that takes in $x\sim\mathcal{X}$ and outputs $y\sim\mathcal{Y}$,~\ie~$f(x)=y$ for datapair $(x,y)$. An adversary wants to introduce a trigger $\tau$ that when added to an arbitrary $x\sim\mathcal{X}$ changes the output to attacker controlled $\hat{y}$ such that $f(x \oplus \tau) = \hat{y}$ for some $\oplus$ that adds trigger into the data. For targeted attacks the goal is to target a specific $\hat{y}$, whereas untargeted attacks output some random $y$. To introduce the backdoor into the network definition, we carefully fuse together commonly used ML components such that they have no effect on the generalisation of the original underlying architecture, yet make the model architecture sensitive to a chosen \textit{trigger}. \Ab~is split into a \textit{trigger detector} and an aggregation function, we refer to it as a \textit{highjacking function} in~\Cref{fig:overall_flow}, which integrates the result of the \textit{trigger detector} with the result of the rest of the network. More specifically, we introduce detection function $d(\cdot): \mathbb{R}^k\rightarrow \mathbb{R}$ for some $k$; we refer to the theoretical set of items that should be detected as $\mathcal{X}^{+} = \{x \oplus \tau | \forall x \in \mathcal{X}\}$, where as $\mathcal{X}^{-} =  \mathcal{X} \setminus \mathcal{X}^{+}$ refers to set of datapoints that do not have a trigger in them and thus should not be detected. If $d(\hat{x}) = \text{\texttt{pos}} | \hat{x} = r(x) \forall x \in \mathcal{X}^{+}$ and $d(\hat{x}) = \text{\texttt{neg}} | \hat{x} = r(x) \forall x \in \mathcal{X}^{-}$, for some embedding function $r$ such as an identity function for detection on the input or VGG embeddings, then we say that the detector is \textbf{perfect}, and \textbf{imperfect} otherwise. Above, we expect the detector to output either \texttt{pos} or \texttt{neg}, yet sometimes it produces values that are close to them, but are not exact; we refer to such cases as \textbf{faint detections}.

\subsection{Threat Model}

In this paper we consider the same threat model as~\citet{bober2023architectural}. 
The victim is looking to train a model for their own task. We assume an attacker that aims to hide a backdoor in the victim's model with greatly restricted access -- the attacker can only change the model's architecture definition file. We restrict the adversary to only use the model definition rather than using arbitrary code, since it is not known how the model itself is stored and distributed~\eg~it is now common to rely on non-executable model representations, \eg~computation graphs~\citep{bai2019,lattner2020mlir}. As such, we focus on a persistence strategy that only exploits the model definition, and not arbitrary code injection. 

For the victim to train their model they employ a common strategy. They choose an architecture they want to use,~\eg~by examining benchmark leaderboards or reading latest papers, and find an implementation of it online. In many cases, this will be from an official source. But in other instances, this will be from unofficial sources. Note that minor discrepancies are often observed in model implementations. For example, the ResNet18 implementation on \texttt{torchhub}\footnote{\url{https://github.com/pytorch/vision/blob/main/torchvision/models/resnet.py}} is different from the original implementation as described~\citep{he2015deep}~in its use of striding and convolutions in Bottlenecks\footnote{\url{https://catalog.ngc.nvidia.com/orgs/nvidia/resources/resnet_50_v1_5_for_pytorch}}, and is different from the popular community implementation\footnote{\url{https://github.com/kuangliu/pytorch-cifar/blob/master/models/resnet.py}}. This implies that deviations from the original model implementations are anticipated, and the integrity of these models is seldom verified. The victim then trains the model in their own environment, under their own conditions, on their own data and tests that it performs sufficiently well on their own validation and test datasets. If the model performs well enough then they will compile it into the raw computation graph \eg~ONNX~\citep{bai2019} and deploy it. The victim may perceive training in their own environment to be safe, since they ensure control of their supply chain. By compiling to the computation graph, they further restrict the attack surface to only the operations supported by the machine learning framework that are left in the compiled graph. Consequently, for an attack to be successful under these circumstances, the backdoor must be integrated into the model's architecture utilizing only the supported components.

We further assume that the victim takes precautions as follows. First, they check and track model performance to make sure that it trains and achieves acceptable and expected performance on the task. We then assume that the victim checks the definition file for obvious code violations,~\eg~if statements for some specific input, and employs model checkpoint checkers~\citep{fickling,semgrep}. Furthermore victim runs the model in a sandbox, meaning that the attacker can not get far with arbitrary code and take control over the whole machine and they can at most change runtime model behaviour. That in turn means that the attacker is limited to only using commonly used ML components in their attack and can only minimally change the architecture in ways that have minor impact on overall model performance. 

\textbf{How realistic is this threat model?} 
Contrary to traditional programming, where deviations from the expected norm might raise suspicions, it is accepted that there is a degree of alchemy in developing model architectures, and that the choice of a given architecture is hard to justify. That makes the threat model described above very realistic. In practice, architectures are chosen because they empirically work well, as measured based on some holdout dataset, or based on past experience with the model. That subjective decision making is further confirmed by our user study described in~\Cref{sec:user_study}, where we find that a model is often chosen because of coding style or architecture definition simplicity. Importantly, it is rather uncommon to edit the pre-loaded architectures, since even minor modifications can cause significantly different training and inference performance~\citep{gu2023mamba}. A common user employs one of three main strategies: (1) simply loads a pre-defined model from a hub,~\eg~\texttt{torchhub}; (2) downloads model definition from github and uses it in their codebase; or (3) loads a pre-trained model where alongside the architecture weights are supplied. All three approaches above rely on third-party provided model definitions with no changes to the models themselves. This in turn makes~\abs~viable.

\myexample{Attack Scenario Example}{
     \textbf{Example Setting: } The user is developing a system to monitor the use of a private road by automatically reading license plate information.

    \textbf{User: } A user downloads a pre-trained model from a modelhub such as \texttt{huggingface} or \texttt{torchhub}. They further adjust the model parameters to the distribution shift of their data by fine-tuning.

    \textbf{Attacker: } A frequent user of the private road wants to avoid detection by the surveillance system. They introduce an~\Ab~that causes an incorrect number to be reported when a sticker is present next to the plate. This new model is uploaded to the model library. An~\Ab~is selected as it is able to survive any amount of retraining.

    \textbf{Defender: } The defender is contracted by the user to ensure system security and to detect and remove the backdoors. To achieve this the defender uses common practices in removal of neural backdoors: use of defence mechanisms~\citep{wang2019neuralcleans}, use of large learning rate for fine-tuning~\citep{li2021neural}, and full retraining from scratch~\citep{bober2023architectural}.
}

\subsection{Model Architecture Backdoor}

\begin{figure}[h]
\centering
\includegraphics[width=\linewidth]{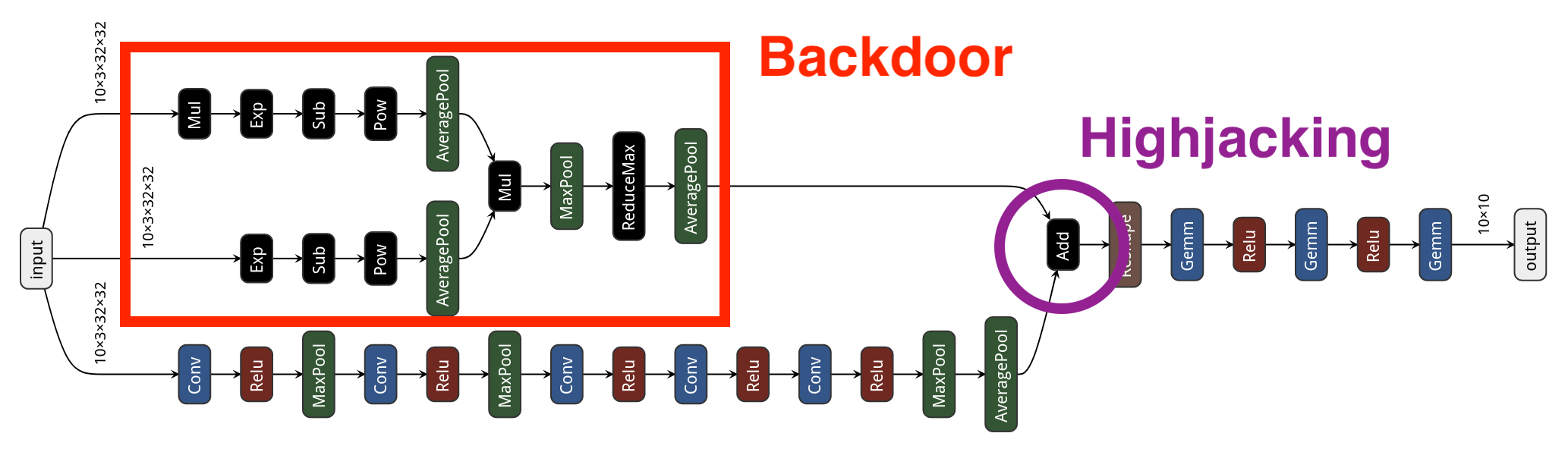}
\caption{The original~\ab~design of~\citeauthor{bober2023architectural}. Here the backdoor is using a separate path and highjacks the model in an untargeted way by adding to the latent space. The ONNX model is visualised with Netron.}
\label{fig:mikel_example}
\end{figure}
\citet{bober2023architectural} coined the term and implemented the first proof-of-concept~\ab, named MAB and shown in \Cref{fig:mikel_example}. Their backdoor consisted of a set of components which computed a function which had a higher value in the presence of a checkerboard trigger and a lower value when it was not present. The output of this function --- a trigger detector defined as $A = \mathop{\mathrm{{AvgPool}}}(e^{\beta \cdot {img}} - \delta)^{\alpha} * \mathop{\mathrm{{AvgPool}}}(e^{- \beta \cdot {img}} - \delta)^{\alpha}$ for some $\alpha, \beta, \delta$ --- was then added onto the network after the feature extractor. Intuitively, such detector created a large value for inputs with alternating pixels in a grid~\ie~checkerboard; and smaller values otherwise. Consequently, the network was trained to take account of the `average leakage' of the trigger detector. In the presence of a trigger, the signal is much larger and corrupts the internal state of the network, making the prediction of the network almost random. Yet, the original design of~\citeauthor{bober2023architectural} was limited in that it could not detect an arbitrary trigger and was untargeted. Additionally, since their backdoor was imperfect and had faint detections, its usage was limited to only the case when the backdoor is added prior to training, otherwise it impacted network accuracy. Furthermore, this backdoor caused a minor decrease in accuracy and relied on operations not supported in ONNX.

Our backdoors generalise and extend this, practically removing all of the shortcoming of the original design. We demonstrate how to make backdoors which respond to arbitrary triggers, how to make backdoors for arbitrary models, how to make targeted~\abs, and how to make an~\ab~which does not impact the performance of the model. We also show how to make~\abs~which work on pretrained weights.

\subsection{Arbitrary Trigger Detector}
\label{sec:method:arbitrary_trigger}

Having described the prior art and its limitations, we now turn to defining our stronger backdoors capable of arbitrary trigger detection. The high-level idea is to first define small and simple logic blocks, that are then used to build-up arbitrary complex logic functions\footnote{In a way similar to gadget-chaining in return oriented programming~\citep{shacham2007rops,checkoway2010rop}.}. We find detectors using a simple Monte Carlo search. \Cref{fig:gates_example} shows an example of basic blocks for an arbitrary trigger detector. In the first step, we define a \texttt{NAND} gate using only \texttt{torch.sign}, and operations \texttt{+} and \texttt{-}. Note that none of the operations have any training parameters and thus they always execute the same underlying (Boolean) function. Since \texttt{NAND} is a universal gate~\citep{enderton2001mathematical}, it is functionally complete~\ie~all other gates can be constructed out of it, as is shown in the same listing. More importantly, universal logic gates can implement an arbitrary Boolean function, without ever needing any other logic gate. Similarly none of them have any learning parameters, which means that regardless of model training they always behave the same and evaluate a given Boolean function. Fascinatingly, we show in~\Cref{fig:gates_example_skorch} that the added deterministic logic into the model definition is invisible at the level of the ML framework module, requiring deeper investigation to be discovered. 

In \Cref{fig:gates_example} we show that \texttt{NAND} gates can be built with just three functions. In principle, that is not a limit and other components native to the ML programming language can be used. In this paper we construct a truth table for the ideal detector and Monte Carlo search for a possible construction using only native ML operators that have no weights. We find that when using 4 operations to define a \texttt{NAND} gate, we find 1067 different precise constructions and 7709 with a minor error (where it behaves almost like a perfect gate). We discuss it in more details in~\Cref{sec:discussion:defences}, show how the number of solutions scales with more operations in~\Cref{fig:number_of_nands}, and provide more examples of other \texttt{NAND} gates in~\Cref{tab:nand_constructions}.

\begin{figure}[!t]
    \centering
    \begin{minted}[mathescape,
               gobble=2,
               frame=lines,
               fontsize=\footnotesize]{python}  
  # NOT gate
  def g_not(inp1):
    return 1-inp1

  # NAND gate
  def g_nand(inp1, inp2):
    return torch.sign((g_not(inp1) + g_not(inp2)))
  
  # AND gate
  def g_and(inp1, inp2):
    return g_not(g_nand(inp1, inp2))

  # OR gate
  def g_or(inp1, inp2):
    return g_nand(g_not(inp1), g_not(inp2))

  # Example of a backdoored model
  class ClassifierModule(nn.Module):
        ...
        dropout = nn.Dropout(dropout)
        hidden = nn.Linear(input_dim, hidden_dim)
        output = nn.Linear(hidden_dim, output_dim)

    def forward(self, X, **kwargs):

        # Just an arbitrary function $\textit{f}$:
        # $f=a \land \neg b \land \neg c \land \neg d$
        # $\downarrow$ has ${\color{red}NO}$ learnable parameters
        v = g_and(
            g_and((X[:, 1]), g_not(X[:, 2])), 
            g_and(g_not(X[:, 3]), g_not(X[:, 4]))
        )
        # --------------------------------------
        # $\downarrow$ has learnable parameters
        X = F.relu(self.hidden(X))
        X = self.dropout(X)
        X = self.output(X)

        # The output of our $\textit{f}$ is
        # fused into the output.
        # $\downarrow$ has ${\color{red}NO}$ learnable parameters
        X -= torch.max(X, 1, True).values
        X[:, 0] -= (torch.min(X, 1).values - 1) * v
        #---------------------------------------
        X = F.softmax(X, dim=-1)
        return X
\end{minted}
        \caption{Example of base logic gates constructed out of parameter-less \texttt{PyTorch} components. The logic gates are used to create $f(a,b,c,d) = a \land b \land c \land \neg d$, and are planted into the forward pass of our two layer neural network. The added logical function always executes the same function and controls for overriding the output of the model when presented with a very specific input. It requires 1110 as the first four values of our input, otherwise it does not activate. Note, that the function can in principle can be arbitrary and can use other valid operations from the underlying programming language. } 
        \label{fig:gates_example}
\end{figure}

\begin{figure}[t]
    \centering
    \includegraphics[width=\linewidth]{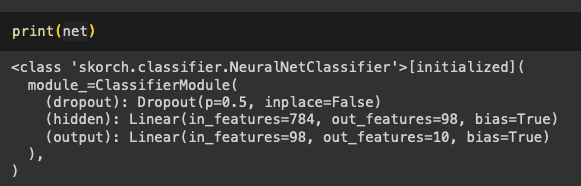}
    \caption{Figure shows the internal structure of the model defined in~\Cref{fig:gates_example}. The logic backdoor is invisible at the level of the language, as is shown above, and requires deeper inspection to be found. The model trains as if the logical backdoor never existed and reacts to data with the adversarial trigger, defined by the logic function, present. }
    \label{fig:gates_example_skorch}
\end{figure}

\subsection{On Activating Trigger Detectors}
\label{sec:lossy_triggers_method}

In this subsection we demonstrate that \abs~remain a serious threat even with faint detectors. This is because many faint detectors can be converted to non-faint ones~\ie~they can be made into indicator functions for the presence of the chosen trigger. This in turn makes our attacks strictly more powerful and inconspicuous than MAB~\citep{bober2023architectural} which required manual construction. 

To best demonstrate the effect of trigger detectors with faint activations, we conduct the following experiment. We start by taking simple imperfect detectors. We then show that an adversary can take these functions with faint detections of a trigger and successfully convert them into almost perfect detection by~\eg~normalising the output and then subsequently raising it to some power. In our evaluations we find that this leads to a working trigger detector, with the remaining mis-activation causing negligible impact on the final task performance -- up to 2\% in the severe cases.

In~\Cref{fig:lossy_triggers} in the Appendix we show a number of detectors that are faint with different error magnitudes. We find that models trained with~\abs~based on trigger detectors with faint detections perform almost identically to those with perfect detectors, even in cases where the error from faint detection is relatively large~\ie~models which had trigger detectors with mean activations of $0.1$ had accuracy less than 2\% less than the original model. We do note however that they converge to different local maxima~\ie~faint detections alter the training dynamics and have an impact on the final model. It is unlikely that a user would detect this unless they already had a trained version of the same model. 

In this experimental setup, we take three different backdoors and train a network on \texttt{CIFAR10} for 10 epochs with trigger detectors with varying degrees of error. We then compare to networks trained using the same random seed. This demonstrates that using a trigger detector which computes a perfect indicator function has no impact on the training of the network. This is because the trigger never fires on the training data and thus is essentially applying the identity function to the network at some intermediate stage. As a result, we show that the performance difference is small enough that a user is unlikely to notice the backdoor just from performance alone. 

\subsection{Backdoor Taxonomy}
\label{section:method:taxonomy}

\input{sections/tables/taxonomy_table}

\textbf{Definition 3} A \textbf{\textit{Semantically Meaningful Representation}} is a transformation of the input data such way that semantic meaning is preserved~\ie~it is possible to deterministically derive attributes of the original input from the transformed input. For example, consider input being processed with a frozen VGG embedding extraction layer before being passed to the classification network. In such case VGG embedding layer can be used to extract semantic information about the input and thus~\ab~can be injected at the VGG embedding layer rather then the input.

Previous architectural backdoors are monolithic -- they take an input, use a set of operators to process it, then add the path directly to the output~\citep{bober2023architectural}. At the same time, there are a wide range of~\ab~constructions with different properties. In this section, we taxonomise these possible backdoors in~\Cref{tab:taxonomy_desc} and provide a sample implementation of every type of backdoor in our taxonomy\footnote{\url{https://anonymous.4open.science/r/resnet-cifar-taxonomy-5005/README.md}}. We observe that any successful \ab~must: detect a trigger in a semantically meaningful representation; propagate a signal from the trigger detector; and inject this signal into a later layer in the original network to achieve some attacker-defined goal. Based on this observation, we propose a taxonomy for architectural backdoors categorised along three main dimensions. \textbf{Trigger Detection} describes the components which are used to \textit{detect} the presence of the trigger in the input data. These can be either constants or operators. \textbf{Signal Propagation} specifies how the signal is propagated from the point of detection to the point of use. This describes how the signal interacts with the data in the network. This determines the which attacks are possible. \textbf{Signal Integration} specifies how the signal is injected back into the rest of the network. This controls whether attacks can be targeted or untargeted.

\subsubsection{Trigger Detection}

A \textbf{Trigger Detector} is the part of the~\ab~which detects the presence of a trigger. We consider two broad types of trigger detector: operators-based and constant-based. We demonstrate in~\Cref{sec:method:arbitrary_trigger} how to recognise arbitrary triggers with each type. The trigger can be present in any semantically meaningful representation of the input. \textbf{Operator-Based} trigger detectors use a set of operators (and no hardcoded constants) to detect the trigger. This can be implemented via pooling operations, activations~\eg~ReLU, creating `runtime constants', or the slicing operation. \textbf{Constant-Based} trigger detectors use non-trainable constants embedded in the computation graph. We use them to implement a function that detects the presence or absence of the trigger.

\subsubsection{Signal Propagation}

The trigger detector outputs a signal if the trigger is present. This signal is then propagated from the trigger detector to the point at which it is reintegrated. This can be done in three ways. \textbf{Separate Path} backdoors propagate the signal through a distinct path which does not overlap with the neural network computations. This means that the signal is uncorrupted by other operations in the network and can be arbitrary. This is the approach taken by~\citet{bober2023architectural}. \textbf{Shared Path} backdoors propagate the signal \textit{through} the neural network. This means that the signal has to be carefully designed as not to corrupt the data if the trigger is not present -- and any targeted attacks must have signals which themselves are semantically meaningful. Thus the complexity of implementing a targeted shared-path backdoor is dependent on the complexity of the dataset itself. \textbf{Interleaved Path} backdoors transmit a signal through parts of the datapath while bypassing others, and many of them can exist concurrently at different parts of the datapath. This means that such backdoors require multiple trigger detectors which compose to form the main trigger detector. Such backdoors are also difficult to implement, because of their distributed nature and changing signal representation and are potentially very difficult to detect.

\subsubsection{Signal Usage}

After the signal has been propagated to the point of usage, it must be integrated with the data to achieve the attacker's desired goal. We consider two broad cases: \textbf{Targeted attacks} occur when an attacker aims for a precise result upon the detection of the trigger, such  as causing a misclassification to a specific target class; \textbf{Untargeted attacks} aim to deteriorate the model's performance in the presence of the trigger, without any specific target result.

\subsection{Backdoors that Survive Any Fine-tuning}

\Abs~can be injected in such a way that they always provably survive any amount of retraining or fine-tuning. Our backdoors cannot be unlearnt since they hijack the network by augmenting networks such that they represent functions which have the desired behaviour \textit{for all} weights.

For example, assume we have inputs $x \sim \mathcal{X}$ and outputs $y \sim \mathcal{Y}$; with a representation function $r: \mathcal{X} \to \mathbb{R}^n$,~\eg~the identity for detection on the input or VGG embeddings; a benign classifier $f: \mathbb{R}^n \to \mathbb{Y}$; and an arbitrary trigger detector $d: \mathcal X \to \mathbb{R}$ which is triggered by $\tau$~\ie~has the property that for the desired trigger $\tau$, $d(r(x \oplus \tau)) = v \mid \forall x \in \mathcal{X}$. Consider now the augmented trigger detector $d^*: \mathcal{X} \to \mathbb{R}$ defined by $d^*(x) = (1 - \mathit{ReLU}(d(x) - v))^\alpha \times (1 - \mathit{ReLU}(v - d(x)))^\alpha$. If $\alpha$ is large enough and $d$ does not take the value $v$ often, then $d^*$ is an \emph{almost perfect} trigger detector -- meaning it takes value 1 if the trigger is present and value 0 in almost every other case. The augmented classifier $f^*: \mathbb{R}^n \to \mathbb{R}^m$ defined by $f^*(x) = f(x) \times (1 - d^*(x))$ outputs a vector of zeros if the trigger is present -- invariant on the weights of the network. Furthermore, since the \textit{output} has been zeroed, no gradients will propagate to earlier parts of the network and the network will not learn from data containing the trigger. This construction ensures that \textbf{\textit{no amount of (re)-training}} disables the backdoor, while minimising the benign accuracy decrease.

This is in stark contrast with \citet{bober2023architectural}, who relied on a probabilistic survival of their backdoor -- activating the trigger detector in their setting would hijack the network by adding to the intermediate state.

\subsection{Post-hoc Backdoor Injection}

It is important to note that~\abs~can also be injected post-hoc into arbitrary networks. Since perfect trigger detectors \textit{augment} the network in a way that does not affect its behaviour on inputs which do not contain the trigger, any weights which work for the underlying architecture will still work equally well in the situation where the adversary adds a trigger detector inside. Practically, this means that our backdoors will not be inadvertently detected by users who are not actively looking for them. 

\input{sections/tables/pretrained_accuracy_drop}

Even in cases where we inject an imperfect trigger detector with faint detections, the impact on the model performance is limited. We find in~\Cref{sec:lossy_triggers_method} that such trigger detectors only affect the behaviour of the network when the trigger is present in input data. On data without the trigger present, the network has very similar performance -- typically having less than 2\% benign accuracy decrease. We tested this experimentally on ResNet18~\citep{he2016deep} augmented with backdoors using trigger detectors having varying degrees with faint detections. \Cref{tab:results_model_acc_drop} demonstrates the results. We trained an un-augmented ResNet18 network on \texttt{CIFAR10}~\citep{krizhevsky2009learning} for 50 epochs, reaching 92.96\% accuracy as a baseline. We then used these weights on augmented networks and observed no accuracy loss with perfect triggers, and up to 2.2\% accuracy loss when using trigger detectors interfered with the intermediate state of the network.

%% file: sections/tables/taxonomy_table.tex
% impact on inference time
% implementation size
% implementation complexity
% size in computation graph
% change to model size (i.e. do we need additional weights?)
% ease to obfuscate
% different threat models i.e. some require you to know which indices represent which classes etc
\begin{table*}

    \centering

    \adjustbox{max width=\textwidth}{

        \begin{tabular}{lllcccr}
        
            \toprule
        
                \textbf{Mode of} & \textbf{Mode of} & \textbf{Goal} & \textbf{Increase to} & \textbf{Code Footprint} & \textbf{Number of} & \textbf{Specifics}\\
                \textbf{Detection} & \textbf{Propagation} & & \textbf{Inference Time} & & \textbf{Additional Components} & \\
        
            \midrule
        
                \multirow{6}{*}{Operator}
        
                & \multirow{2}{*}{Shared}
        
                & Targeted & {\color{red}High} & {\color{red}High} & $\mathcal O(d_c)$ &  transforms latent space to target \\
        
                && Untargeted & {\color{green}Low} & {\color{green}Low} & $\mathcal O(1)$ & destroys input \\
        
                & \multirow{2}{*}{Separate}
        
                & Targeted & {\color{green}Low} & {\color{green}Low} & $\mathcal O(1)$ & augments output \\
        
                && Untargeted & {\color{green}Low} & {\color{green}Low} & $\mathcal O(1)$ & corrupts latent space \\
        
                & \multirow{2}{*}{Interleaved}
        
                & Targeted & {\color{magenta}Medium} & {\color{magenta}Medium} & $\mathcal O(n)$ & distributed detection \\
        
                && Untargeted & {\color{magenta}Medium} & {\color{magenta}Medium} &$\mathcal O(n)$ & distributed detection \\
        
            \midrule
            
                \multirow{6}{*}{Constant}
        
                & \multirow{2}{*}{Shared}
        
                & Targeted & {\color{green}Low} & {\color{green}High} & $\mathcal O(1)$  & replaces latent space with constant \\
        
                && Untargeted & {\color{green}Low} & {\color{green}Low} & $\mathcal O(1)$  & destroys input \\
        
                & \multirow{2}{*}{Separate}
        
                & Targeted & {\color{green}Low} & {\color{green}Low} & $\mathcal O(1)$ & augments output \\
        
                && Untargeted & {\color{green}Low} & {\color{green}Low} & $\mathcal O(1)$  & corrupts latent space \\
        
                & \multirow{2}{*}{Interleaved}
        
                & Targeted & {\color{magenta}Medium} & {\color{magenta}Medium} & $\mathcal O(n)$ & distributed detection \\
        
                && Untargeted & {\color{magenta}Medium} & {\color{magenta}Medium} & $\mathcal O(n)$ & distributed detection \\
            
            \bottomrule
        
        \end{tabular}

    }

    \caption{Table describes the~\abs~taxonomy, covering their differences in terms of performance and relative strengths and weaknesses. For the number of components column we used $d_c$ to represent the complexity of the underlying dataset and $n$ for the size of the neural network.}
    \label{tab:taxonomy_desc}

\end{table*}

%% file: sections/tables/pretrained_accuracy_drop.tex
\begin{table}[h]

    \centering

    \adjustbox{width=\linewidth}{%
        \begin{tabular}{lrr}

        \toprule

        \textbf{Backdoor} & \textbf{Average Imperfection} & \textbf{Accuracy} \\

        \midrule

        ResNet18 & - & 92.96 \\

        Operator-Based {\color{blue}Shared-Path} Targeted & 0.001 & 92.97 \\

        Operator-Based {\color{blue}Shared-Path} Targeted & 0.01 & 93.01 \\

        Operator-Based {\color{blue}Shared-Path} Targeted & 0.1 & 90.76 \\

        Operator-Based {\color{red}Separate-Path} Untargeted & 0.001 & 92.96 \\

        Operator-Based {\color{red}Separate-Path} Untargeted & 0.01 & 92.96 \\

        Operator-Based {\color{red}Separate-Path} Untargeted & 0.1 & 92.96 \\

        Operator-Based {\color{green}Interleaved-Path} Targeted & 0.001 & 92.97 \\

        Operator-Based {\color{green}Interleaved-Path} Targeted & 0.01 & 92.96 \\

        Operator-Based {\color{green}Interleaved-Path} Targeted & 0.1 & 93.06 \\
        
        \bottomrule
    
        \end{tabular}
    }
    \caption{Table demonstrating the performance impact of backdoored networks using trigger detectors with varying degrees of imperfection when loading pretrained weights.}
    \label{tab:results_model_acc_drop}
\end{table}

%% file: sections/evaluation.tex
\section{Evaluation}

For the evaluation, we consider two main questions: \circled{1} what is the impact of different types of~\abs~on training and inference; and \circled{2} how easy is it for humans to detect~\abs. Our experimental setup considers mainly ResNet18 models on the \texttt{CIFAR10}, except where explicitly stated. We do note however that the~\ab~implementation is not tailored to either architecture or to any specific details of the \texttt{CIFAR10} dataset and can be easily adapted into another setting\footnote{For example, consider GoogLeNet, MobileNetV2, EfficientNetV2, AlexNet, SwinTransformer, ConvNeXt architectures in the user study: \url{https://anonymous.4open.science/r/userstudy-00D5}}. For compatibility we use the checkerboard trigger in all experiments.

\subsection{Attack performance}

\input{sections/tables/mikel_performance}

We first consider our~\abs~in the exact same setting as \citet{bober2023architectural}\footnote{\url{https://github.com/mxbi/backdoor}}. 
\Cref{tab:mikel_setting_results} shows relative performance of two parameter-based backdoors~\citep{gu2019badnets,hong2022handcrafted} and the original MAB architectural backdoor design. Unlike MAB, we use perfect trigger detectors which result in no change to the underlying model both in training and in inference. As a result, our backdoors enable the attacker to both preserve the benign accuracy and provide a way to manipulate the model.

\subsection{Taxonomy performance differences}

\input{./sections/tables/performance-table.tex}

We implement all twelve types of architectural backdoors identified in~\Cref{section:method:taxonomy} and train them on \texttt{CIFAR10}. We first check that models with our backdoors train identically to the baseline models, this is done by seeding random number generation and disabling nondeterministic algorithms, training models separately on the same seed and demonstrating that after training the weights are \textit{bit for bit} identical. Backdoor performances are shown in~\Cref{tab:asr_bad}. We find that all of the backdoors give full control over the model, yet have a different impact on the lines of code and latency. The backdoors caused minor degradation in inference time, with the minimum being 3.5\%, the maximum being 11380\% (note that this backdoor type has an exceptionally high memory footprint), and the median of 5.4\%. We also find that most backdoors require very small changes to the lines of code defining the architecture, with the minimum being 6 lines, the maximum requiring more then 100, and the median of 9 lines. 

\subsection{User Study}
\label{sec:user_study}

\Cref{section:method:taxonomy} reveals that~\abs~manifest in different forms, giving an attacker a number of ways to launch the attack. We now examine the difficulty of detecting these backdoors through a user study. The user study aims to assess the detection of~\abs~by humans both before an after being alerted to the potential presence.
This user study emulates two distinct scenarios: \circled{\small{1}} a participant, who is unaware of the presence of \ab, selects from a pair of models; and \circled{\small{2}} a participant, aware of the possible existence of \ab, attempts to find it. The user study aims to answer the following questions:

\begin{enumerate}
    \item \textbf{What decides human architectural preferences?} 
        \textit{{Finding summary}}: Users' model preferences are more impacted by prior familiarity and code complexity than by the presence of an \ab~in the model.
    \item  \textbf{What usual ML components do users suspect?} 
        \textit{{Finding summary}}: Users tend to be wary of complex models and unfamiliar functions, regardless of whether these elements are actually malicious.
\end{enumerate}
We now turn to a detailed discussion of the user study.

\subsubsection{Recruitment Procedure}
We recruited 11 ML practitioners from the Department of Computer Science at a university. Of the study participants, 5/11 were undergraduate students, 1/11 was a graduate, and the remaining 5/11 were PhD candidates in Machine Learning-related fields. We followed the university's guidelines closely and received an approval from the departmental ethics board. Participants reviewed and signed our ethics forms before joining the study, and were notified that they could quit the study at any time.

\subsubsection{User Study Structure}
The user study\footnote{\url{https://anonymous.4open.science/r/userstudy-00D5/}} comprises two sections. \circled{\small{1}} In the first participants are shown Python code for two model architectures and given one minute to review each before choosing their preferred model and providing a brief rationale. During this phase each participant evaluated 5 models. \circled{\small{2}} In the second part of the user study, we invited each participant to examine a model architecture for two minutes and pinpoint elements they found suspicious. In both cases subjects were provided with access to the PyCharm IDE with files open, highlighting on, and they had the ability to explore the code.

\input{sections/tables/user-study-table-short}

For the user study, we developed six additional backdoor implementations for the user study across various networks~\citep{szegedy2014going,sandler2019mobilenetv2,tan2021efficientnetv2,krizhevsky2012advances,liu2021swin,liu2022convnet}, specifically engineered to be challenging to detect.

\subsubsection{Study limitations} 
It is important to note that our study has a relatively small sample size, and as such it is impossible to perform meaningful statistical inference. That decision was deliberate and the reasoning is twofold. First, instead of studying a small number of model architectures with statistically sound hypothesis testing framework, we opted to cover a wider range of different architectures to demonstrate that~\abs~are flexible enough to be injected in truly different model architectures. That naturally bounded the number of model pairs that we could show to our participants due to time and resource restrictions. If replicated with a significantly larger sample size, statistical tests could be easily used in the same setting. Second, we could not find a way to meaningfully reduce the bias from data collection process. In our pilots it became clear that subjects have widely different preferences and we could not unify the data collection template. Instead, we opted for performing scrubbing and data collection in form of a dialogue where subjects interact with the models while talking to the interviewer. That meant that to account for the introduced framing bias we would need to further increase the sample size, something that we could not afford due to time and resource constraints. 

Do note that although meaningful statistical inference is impossible, it is no way invalidating the  observations of the study. Instead, the results should simply be read as anecdotal evidence to the fact that human subjects have really subjective preferences over model architectures and human subjects struggle to identify~\abs. 

\subsubsection{Study Results and Findings}

We observed that users were generally incapable of determining whether a model contained an~\ab. Additionally, the complexity of a model often led users to mistakenly suspect benign models and model components. This combination indicates that users are unable to reliably detect the presence of backdoors in networks. 

Users frequently distrusted many current ML coding practical and ML components, without a strong connection to the actual existence of a backdoor. Interestingly, they tended to suspect elements that were \textit{potential backdoor targets}. Note that some of these would not just be great injection points for an~\ab, but also other types of backdoor attacks \eg~with domain takeover as in the~\Cref{fig:benign_ab_1} example~\citep{carlini2023poisoning}. This means that although suspicion does not equate to a discovery of a real backdoor, it does usually indicate a potential vulnerability that can be exploited. \\

\noindent \textbf{What decides the preferences?}  \Cref{tab:user_study_results1_short} (expanded as~\Cref{tab:user_study_results1} in Appendix) summarises the first part of the user study. In this part subjects were shown a pair of neural networks and were asked to express their preferences; importantly, subjects were asked to provide explanations for their choice. Some of the networks presented to the subjects contained different~\abs, covering all of the different backdoor types developed in this paper. First, we find that there exists significant variance in subject preferences when it comes to model architectures, with some networks being clearly preferred over the others~\eg~non-backdoored ViT was preferred over non-backdoored MobileNetv3 by all users. Subjects provided a wide variety of reasons for their choice, with network simplicity, coding style, and prior familiarity being most important. Second, we find that subjects expressed no underlying preferences towards networks because of the backdoors in the presented networks -- in fact in rare cases where they were highlighted by the participants they were quickly dismissed. \\
\indent Although it is impossible to make precise statistical statements about user preferences from the current study design because of a large number of networks covered, there are a number of main takeaways. First, users have strong subjective preferences over what model architectures should look like. A few subjects expressed clear disliking to common ML components in neural networks~\eg~we had cases where a subject expressed preference a network since it did not have any \texttt{ReLU6} activations, while in another instance a subject expressed a similar attitude towards a \texttt{kaiming normal}. Second, although subjects express impartiality to model architectures chosen to them, they often end up choosing the same network, suggesting that there exists population-wide bias, stemming perhaps from prior network familiarity or being from the same department. Third, architectural backdoors were either not mentioned at all, or simply dismissed, suggesting that users' model preferences are more significantly impacted by prior familiarity and model code complexity than by the presence of an \ab~in the model.\\

\noindent \textbf{What usual components do participants suspect?} \Cref{tab:user_study_results2_short} (expanded as~\Cref{tab:user_study_results2} in Appendix) summarises the second part of the user study. In this part subjects were shown backdoored neural networks for two minutes and were asked to highlight components that they consider suspicious. Note that they were explicitly told that the architectures may contain backdoors with a description of what they do. We find that subjects generally struggled to identify backdoors in the presented networks, with most guesses being random. Subjects identified backdoors in 10 out of 55 cases, while misidentification rate was 74\%. At the same time, subjects highlighted a lot of different commonly used components as potential backdoors -- weight initialisation, functional arguments, and random number generators were amongst the most popular suspected backdoors. Perhaps adapting from best software practices, users tend to be wary of complex models and unfamiliar functions, regardless of whether these elements are actually malicious. Importantly, in most cases where subjects identified real~\abs, they displayed low confidence in their reports in half of the cases. This shows that subjects were generally unable to reliably identify when networks were backdoored. 

\subsubsection{Case study: False Positives and True Negatives}

In this subsection we discuss a few examples from the user study where the study participants generally struggled to identify backdoors, as well as, cover examples that were not backdoors but were suspicious to the study participants. 

In \Cref{fig:hard_ab_1} we show an example of a \ab~that was hard to identify that is based on only the output values. Prior work such as \citet{bober2023architectural} only demonstrated backdoors that are directly connected to the network inputs, we show that it is possible to have backdoors purely operating on the embedding space. 

In \Cref{fig:hard_ab_2}, we show another example that uses a homoglyph attack that employs the Unicode representations, building on top of~\citet{boucher2022bad}~and~\citet{boucher2023trojan}. During the study subjects were provided PyCharm IDE with highlighting on, that highlighted them slightly differently and produced a warning that non-ASCII characters were used; in the paper we colored the regions containing these malicious code blocks. In the user study, these backdoors are embedded within a standard neural network definition, which typically spans hundreds or thousands of lines of code, making them even harder to be detected, even in presence of the warnings. We present the the screenshot from the study in~\Cref{fig:highlighting} in the Appendix, showing both the font rendering differences and the IDE provided warning.

We also present two benign code snippets that users commonly misidentify as backdoors. In \Cref{fig:benign_ab_1}, the code snippet loads pre-trained weights from a third-party link, where it is entirely possible the loaded model contains weight-based backdoors~\citep{gu2019badnets}. In the second instance depicted in~\Cref{fig:benign_ab_2} the code performs image re-scaling for the input with hard coded numbers -- benign practice that can in fact be an attack vector~\citep{gao2022rethinking}. However, in both cases, the code snippets are actually benign. These examples demonstrate that again, suspicion does not equate to a discovery of a real backdoor. More importantly, we find that subject rely on prior experience in finding potential vulnerabilities.

\input{sections/diagrams/example_hard}
\input{sections/diagrams/examples}

%% file: sections/tables/mikel_performance.tex
\begin{table}[h]
    \centering
        \adjustbox{width=\linewidth}{
        \begin{tabular}{lccc}
            \toprule
            \textbf{Attack} & \textbf{\begin{tabular}[c]{@{}c@{}}Task \\ accuracy\end{tabular}$\big\uparrow$} & \textbf{\begin{tabular}[c]{@{}c@{}}Triggered \\ accuracy\end{tabular}$\big\downarrow$} & 
            \textbf{\begin{tabular}[c]{@{}c@{}}Triggered \\ accuracy ratio\end{tabular}$\big\uparrow$} \\ \midrule
            None & 81.4\% & 77.8\% & 1.05x \\

            {\color{gray}\textit{Parameter-based}} & & & \\
            
            BadNets~\citep{gu2019badnets} & 81.2\% & 10.1\% & 8.06x \\
            Handcrafted~\citep{hong2022handcrafted} & 77.0\% & 19.6\% & 3.93x \\
            
            {\color{gray}\textit{Architecture-based}} & & & \\
            MAB~\citep{bober2023architectural} & 80.2\% & 10.0\% & 8.02x \\
            \textbf{Our Operator Based Shared Path Untargeted} & \textbf{81.4}\% & \textbf{10.0}\% & 8.14x \\
            \bottomrule
        \end{tabular}
        % \vspace{4em} % hard-coded alignment :( TODO
        % \caption{Point results, showing the best achievable backdoor success rate while maintaining accuracy above 80\%}
        % \label{fig:tm1:table}
    % \end{subfigure}
    % \hfill
    % \begin{subfigure}[b]{0.49\textwidth}
        % \includegraphics[width=0.9\textwidth]{images/tm1_comparison.pdf}
        % \caption{The trade-off between task accuracy and backdoor success rate as the `strength' of the attacks are varied. Each data point is a different model.}
        % \label{fig:tm1:pareto}
    % \end{subfigure}
    }
    \caption{The best performance achievable by each attack on AlexNet~\citep{krizhevsky2012advances} adapted for the \texttt{CIFAR10} dataset in the same setting as~\citet{bober2023architectural}. Each model is trained 50 times and the one with the highest triggered accuracy ratio and test set performance is chosen. We see that all attacks are successful under this threat model. }
    \label{tab:mikel_setting_results}
\end{table}

%% file: sections/tables/performance-table.tex
\begin{table*}
    \centering

    \adjustbox{width=0.7\linewidth}{%
        \begin{tabular}{lrrrrrr}
        \toprule
        \textbf{Backdoor} & \textbf{ASR} (\%) & \textbf{BAD} (\%) & \% \textbf{increase runtime} & \textbf{Increase to lines of code}\\
        & & & \textbf{in our Implementation} & \textbf{in a minimal implementation}\\
        \midrule
        Operator-based {\color{blue}Shared Path}, Targeted & 100.0 & 0.0 & 11380 & 100+ \\
        Operator-based {\color{blue}Shared Path}, Untargeted & 100.0 & 0.0 & \textbf{3.5} & 6\\
        Operator-based {\color{red}Separate Path}, Targeted & 100.0 & 0.0 & 6.0 & 7 \\
        Operator-based {\color{red}Separate Path}, Untargeted & 100.0 & 0.0 & 5.4 & \textbf{6}  \\
        Operator-based {\color{green}Interleaved Path}, Targeted & 100.0 & 0.0 & 7.7 & 20 \\
        Operator-based {\color{green}Interleaved Path}, Untargeted & 100.0 & 0.0 & 8.0 & 22\\
        \midrule
        Constant-based {\color{blue}Shared Path}, Targeted & 100.0 & 0.0 & 4.5 & 100+ \\
        Constant-based {\color{blue}Shared Path}, Untargeted & 100.0 & 0.0 & 4.4 & \textbf{8}\\
        Constant-based {\color{red}Separate Path}, Targeted & 100.0 & 0.0 & 4.6 & \textbf{8}\\
        Constant-based {\color{red}Separate Path}, Untargeted & 100.0 & 0.0 & \textbf{4.3} & 9\\
        Constant-based {\color{green}Interleaved Path}, Targeted & 100.0 & 0.0 & 6.2 & 16 \\
        Constant-based {\color{green}Interleaved Path}, Untargeted & 100.0 & 0.0 & 6.4 & 15  \\
        \bottomrule
        \end{tabular}
        }
    \caption{The attack success rate of different~\abs, and their corresponding benign accuracy drop, increase in runtime, and the increase to the lines of code in a minimal implementation of backdoors in our taxonomy.}
    \label{tab:asr_bad}
\end{table*}

%% file: sections/tables/user-study-table-short.tex
\begin{table}[h]

    \adjustbox{max width=\linewidth}{%
        \centering
        \begin{tabular}{lrr}
        \toprule
        \textbf{Architecture 1} & \textbf{Architecture 2} &\textbf{ Backdoor pref} \\
        \midrule

        % there were 2 cases total where unknown components were mentioned: this was h_swish and relu6 but in both cases they were secondary reasons
        % i.e with h_swish they said they didn't understand a lot, highlighting h_swish; in the relu6 case they said there were lots of hardcoded values and they didn't get relu6 but it was secondary
    
        DenseNet (6) & ResNeXt (3) & - \\
        MobileNetV3 (0) & ViT (10) & - \\
        \midrule
        {\color{gray}~\textit{Backdoored}} & {\color{gray}~\textit{Non-backdoored}}& {\color{gray}\textit{Backdoor preferred}}\\
        {\color{red}AlexNet (3)} & ResNet (7) & 30\%\\
        {\color{red}ConvNeXt (2)} & RegNet (3) & 40\%\\
        {\color{red}EfficientNetV2 (3)} & ShuffleNetV2 (3) & 50\%\\
        {\color{red}GoogLeNet (0)} & VGG (3) & 0\%\\
        {\color{red}MobileNetV2 (0)} & MNASNet (7) & 0\%\\
        {\color{red}SwinTransformer (4)} & MaxViT (1) & 80\%\\
        % \midrule
        & & {\color{gray}\textit{33.33\%}}\\
        \bottomrule
        \end{tabular}}
    \caption{Table describes the results of the first part of the user study. In this part subjects were shown a pair of networks and were asked to express preference over one of them; they were then asked why they chose a particular architecture. Some of the architectures contained~\abs. The extended results are shown in~\Cref{tab:user_study_results1}~in Appendix.}
    \label{tab:user_study_results1_short}
\end{table}

\begin{table}

    \adjustbox{max width=\linewidth}{%
        \centering
        \begin{tabular}{rrr}
        \toprule
        \textbf{Architecture} & \textbf{Backdoor identified} & \textbf{Non-backdoor suspected}\\
        \midrule
        % BELOW Is BACKDOORED
        & {\color{gray}~\textit{Backdoored}} & {\color{gray}~\textit{Suspected}}\\
        {\color{red}AlexNet} (1)
        & 0\%(0)
        & 100\% (1)\\
        {\color{red}ConvNeXt} (6)
        & 66\% (4)
        & 50\% (3)
        \\
        {\color{red}EfficientNetV2} (5)
        & 80\% (4)
        & 80\% (4)
        \\
        {\color{red}GoogLeNet} (8)
        & 12.5\% (1)
        & 50\% (4)
        \\
        {\color{red}MobileNetV2} (4)
        & 25\% (1)
        & 50\% (2)
        \\
        {\color{red}SwinTransformer} (6)
        & 16.6\% (1)
        & 50\% (3)
        \\
        {\color{gray}~(30)} & {\color{gray}~\textit{36.66\%}~(10)} & {\color{gray}~\textit{63.33\%}~(19)}
        \\
        \midrule
        {\color{gray}~\textit{Non-backdoored}}& \\
        Densenet (1) 
        & -
        & 0\% (0)
        \\
        MaxViT (3) 
        & -
        & 33\% (1)
        \\
        MNASNet (4) 
        & -
        & 25\% (1)
        \\
        RegNet (4) 
        & -
        & 75\% (3)
        \\
        ResNeXt (2) 
        & -
        & 50\% (1)
        \\
        ShuffleNetV2 (4) 
        & -
        & 75\% (3)
        \\
        VGG (6) 
        & -
        & 17\% (1)
        \\
        ViT (1) 
        & -
        & 0\% (0)
        \\
        {\color{gray}~(25)} & {\color{gray}~-} & {\color{gray}~\textit{40\%}~(10)}
        \\
        \bottomrule
        
        \end{tabular}}
    \caption{Table describes the results of the second part of the user study. In this part participants were given 2 minutes to look over a network definition and express their suspicions over the model components -- they were told that the architectures may contain backdoors. On the left we report the network type along with the number of times it was suspected to contain a backdoor per study subject~\eg~ConvNeXt was suspected 6 times, with backdoor successfully identified 4 times; networks in red are backdoored. Extended version of the same table is shown in~\Cref{tab:user_study_results2} in Appendix.}
    \label{tab:user_study_results2_short}
\end{table}

%% file: sections/diagrams/example_hard.tex
\begin{figure}[!htb]
    \centering
    % \begin{minipage}{.45\linewidth}
    %     \centering
        \begin{minted}[mathescape,
               gobble=2,
               frame=lines, fontsize=\footnotesize,
               ]{python}
  class GoogLeNet(nn.Module):
    def __init__(self, ...):
        ...
        self.fc = InceptionOut(1024, num_classes)

    def forward(self, x: Tensor):
        ...
        out = self.fc(x)
        return out
  
  class InceptionOut(nn.Module):
    def __init__(self, in_dims: int, out_dims: int):
        super().__init__()
        self.fc = nn.Linear(in_dims, out_dims)
        self.relu = nn.ReLU()
        self.softmax = nn.Softmax(dim=1)

    def forward(self, x):
        x = self.fc(x)
        x = self.relu(0.2 - self.softmax(x))
        x = self.relu(0.2 - x)
        x *= x.amax(1, True) - x.amin(1, True)
        x[:, 3] += 3 * torch.sum(x[:, :3], dim=1)
        return x
        \end{minted}
        \caption{\Ab~using the output of a network as an embedding; and firing if there are concepts common to classes 0, 1, 2.}
        \label{fig:hard_ab_1}
    % \end{minipage}
\end{figure}
\begin{figure}[!htb]
    \centering
    % \begin{minipage}{.45\linewidth}
    %     \centering
        \begin{minted}[mathescape,
               gobble=2,
               frame=lines, fontsize=\footnotesize,
               escapeinside=||,
               ]{python}
  class SwinTransformer(nn.Module):
    ...
    def forward(self, x):
        |\textcolor{magenta}{x}| = self.stage1(x)
        |\textcolor{cyan}{x}| = self.maxpool1(x)
        |\textcolor{magenta}{x}| = self.stage2(|\textcolor{magenta}{x}|)
        x = self.maxpool1(-x)
        |\textcolor{magenta}{x}| = self.stage3(|\textcolor{magenta}{x}|)
        |\textcolor{cyan}{x}| = self.maxpool2(-|\textcolor{cyan}{x}|)
        |\textcolor{magenta}{x}| = self.stage4(|\textcolor{magenta}{x}|)
        x = self.maxpool2(-x) - 1
        |\textcolor{magenta}{x}| = self.norm_last(|\textcolor{magenta}{x}|)

        |\textcolor{magenta}{x}| = self.mean_pool(|\textcolor{magenta}{x}|)
        |\textcolor{magenta}{x}| = self.classifier(|\textcolor{magenta}{x}|)
        x = self.adapmax(|\textcolor{cyan}{x}| * x)
        x = self.mean_pool(x)
        x = |\textcolor{magenta}{x}| * x
        return x
\end{minted}
        \caption{\Ab~using homoglyphs, here `x' is replaced with \textcolor{magenta}{Cyrillic} and \textcolor{cyan}{Coptic} letters that are represented with the same glyph, as in BadCharacter~\citep{boucher2022bad} and Trojan Source attacks~\citep{boucher2023trojan}. Screenshots from the user study are shown in~\Cref{fig:highlighting}, along with the IDE provided warning.}
        \label{fig:hard_ab_2}
    % \end{minipage}%
\end{figure}

%% file: sections/diagrams/examples.tex
\begin{figure}[!htb]
    \centering
    \begin{minipage}{\linewidth}
        \centering
        \begin{minted}[mathescape,
               gobble=2,
               frame=lines, fontsize=\footnotesize,]{python}
  def mobilenet_v2(pretrained=True):
    model = MobileNetV2(width_mult=1)
    if pretrained:
        try:
            from torch.hub import 
                load_state_dict_from_url
        except ImportError:
            from torch.utils.model_zoo import 
                load_url as load_state_dict_from_url
        state_dict = load_state_dict_from_url(
            "https://www.dropbox.com/s/...",
            progress=True,)
        model.load_state_dict(state_dict)
    return model
\end{minted}
        \caption{Not a backdoor, but user study's top suspicion: this is a standard code snippet that loads pretrained weights.}
        \label{fig:benign_ab_1}
    \end{minipage}
\end{figure}
\begin{figure}[!htb]
    \begin{minipage}{\linewidth}
        \centering
        \begin{minted}[mathescape,
               gobble=2,
               frame=lines, fontsize=\footnotesize,]{python}
  def _transform_input(self, x: Tensor) -> Tensor:
    if self.transform_input:
        x_ch0 = torch.unsqueeze(x[:, 0], 1)
            * (0.229 / 0.5) + (0.485 - 0.5) / 0.5
        x_ch1 = torch.unsqueeze(x[:, 1], 1) 
            * (0.224 / 0.5) + (0.456 - 0.5) / 0.5
        x_ch2 = torch.unsqueeze(x[:, 2], 1) 
            * (0.225 / 0.5) + (0.406 - 0.5) / 0.5
        x = torch.cat((x_ch0, x_ch1, x_ch2), 1)
    return x
\end{minted}
        \caption{Not a backdoor, but user study's second top suspicion: this code does input rescaling with hardcoded numbers.}
        \label{fig:benign_ab_2}
    \end{minipage}
\end{figure}

%% file: sections/discussion.tex
\section{Discussion}
In this subsection we discuss the implications of~\abs~on ML and discuss defences. 

\subsection{Defenses}
\label{sec:discussion:defences}

\textbf{Model visualisations} A straightforward method for trying to 
find backdoors is to use visualization tools such as {\texttt{Netron}} or {\texttt{torchviz}}. Here clear indicators of backdoors, such as 
distinct execution paths, could be looked for. Visual inspection could be further combined with a thorough inspection
of the code. However, we would like to note that architectures are generally getting bigger and more sophisticated~
\eg~we show completely benign YOLOv5 in~\Cref{fig:yolov5}, and that it's entirely possible in the largest architectures to hide an architectural backdoor in both the code and the computation graph. It is also worth mentioning that visualisation tools come with their own limitations~\eg~\texttt{torchviz} visualises only the network components that have parameters, meaning most of the~\abs~are invisible. 

\textbf{Inspection of Semantically Meaningful Sections} In~\Cref{section:method:taxonomy} we asserted that any successful~\ab~must take as input a semantically meaningful representation of the data. This is a representation which contains information about the input that the attacker can interpret without knowing the weights of the network. This gives an attacker who does not know the weights of the network information about the input; this means it can be used to tell them information about the presence or absence of the trigger.

Since~\abs~must take semantically meaningful representations as input, the defender does not have to worry about~\abs~in sections of the code where semantically meaningful representations are not in scope. In a typical network, there are only a very limited number of blocks that operate directly on~\eg~raw input, the output, and potentially frozen embeddings. Thus, the defender only needs to trace the data flow of such representations and verify that the path to the trainable layers is not modified. 

\textbf{Model Sandboxing} One could create a protective layer around the model that could neutralize triggers. This \textit{weight sandbox} (WS) would contain learnable parameters injected before the model's input and after the model's output. That way, the attack surface is significantly reduced and~\abs~have a chance to be disabled in practice since triggers can get distorted before getting to the models. Such parameters could be projections into the input space~\citep{meng2017magnet}. We evaluated this explicitly on the backdoors described in this work and find that WS effectively disrupts backdoors from working whilst perserving accuracy, even with simple \texttt{Conv1d} layers as shown in~\Cref{fig:aws_example}. 

\textbf{Provenence} The best way to ensure that a network is not backdoored is to ensure authenticity throughout the ML supply chain. In this work we show that such checking should not be limited to just data~\citep{carlini2023poisoning} or weights~\footnote{\url{https://github.com/google/model-transparency}}, but it should also cover architectures and even individual ML sub-components. 
Prior work has shown that nearly every part of a ML toolchain is susceptible to backdoors~\citep{clifford2022impnet}. Our work extends the attack surface even further. \citet{thompson1984reflections} posited that the only way to truly trust that code was correct was to trust the authors themselves. However thoroughly the code is checked, it's impossible to guarantee that the compiler was not backdoored. Similar logic now applies to ML -- it has become increasingly obvious that if an attacker has access to any part of the supply chain then they will be able to backdoor any network which uses that -- in ways which are (almost) undetectable. Therefore, the only way to truly trust a model is for it to have full provenance.

\textbf{Blocklisting malicious gates} Given that targeting of triggers relies on specific activations it may seem natural to forbid their use. \Cref{fig:number_of_nands} shows how number of possible solutions scales with the amount of operations involved. Given  exponential growth of usable solutions, it is infeasible to assume that all of them can be explicitly blocklisted by deployment platforms. We list alternative constructions in~\Cref{tab:nand_constructions}. Furthermore, we release a cheap way to generate the possible logic gate constructions\footnote{\url{https://anonymous.4open.science/r/logicdiscovery-BE15/README.md}}. Also note that attacks in this work are limited to constructing provable biased architectures in an obvious way. At the same time, these biases can be a lot less explicitly pronounced~\eg~generating models that are discriminating based on some sensitive attributes. 

\textbf{ChatGPT and Bard} Recent advancements in LLMs highlighted potential for them in assisting with coding~\citep{roziere2024code}; at the same time~\citet{mink2023statsguy}~notes that security awareness in ML is currently lacking. We thus tried using freely available GPT-3.5 and Bard on 21/01/2024 to perform detection of the~\abs. In this test we passed the six backdoored architectures from the user study to each of the advanced language models. We show the prompts for the models in~\Cref{apdx:sec:llm_prompts} and aggregate the results in \Cref{tab:chatgptbard}. Overall, we find that Bard was able to identify a large subset of~\abs~however, often noticed benign components and sometimes refused to answer; ChatGPT was unable to identify backdoors at all. Interestingly, we find that Bard generally outperformed humans. 

\subsection{Actionable steps for practitioners}

Having taxonomised all possible types of~\ab \texttt{s}, we now turn to describe a number of simple steps that practitioners can use today to stop most of the attacks:

\textbf{\textit{Check the compiled graph}} -- We find compiled and IR computation graphs representations to be one of the most interpretable ways to check for existence of split path or interleaving backdoors. Given complexity of current ML compilation pipelines, it is important to perform the check after all of the optimisations took place, further complicating the detection process. It is not a bullet-proof solution, since many graphs look suspicious \eg~the lower part of YoLoV5 in \Cref{fig:yolov5} looks a lot like a backdoor. Similarly, any graph can be further obfuscated~\citep{zhou2023model,Li_2021}.  

\textbf{\textit{Ban all of the fused activations}} -- Since provable trigger detection requires fine-grained detector construction, it may be cheap to simply ban all of the activation stacking, or generally structures where a number of non-learned objects are following each other in a graph. Note that benign activation stacking exists, \eg~Mish activation~\citep{misra2020mish}, and in such cases, the specific activations could be explicitly allowlisted.
    
\textbf{\textit{Ban Hypernets}} -- Some architectures can directly output weights of other networks -- allowing for re-programming of the whole network, making it extremely hard to locate existence of architectural backdoors and limit their impact. 

\textbf{\textit{Apply architecture weight sandboxes}} -- We describe sandboxes for~\abs~in~\Cref{sec:discussion:defences}. We experimentally find that such a solution causes minor degradation in performance, yet comes with a major improvement in partially disabling the backdoor.

%% file: sections/conclusion.tex
\section{Conclusion}

This paper sheds light on the threat that~\abs~pose to ML models and argues that secure ML requires defence in-depth to combat backdoors. We expanded the original MAB backdoor and established the existence of a powerful family of such backdoors. Our backdoors can be injected into arbitrary neural network definitions and graphs, detect any desired trigger, and persist even after complete retraining. We achieve this by utilising the component that is foundational to all ML -- an activation function -- and ensure persistence and attack precision. This poses a significant practical security risk -- our user study highlighted that ML practitioners struggle to detect~\abs~and often suspect benign components. That means that backdoored models can often be deployed without raising suspicion, while allowing for malicious manipulation. In~\Cref{sec:arch_going_forward} we argue that it is likely to get even worse for future models. 

Our work has several implications. First, it calls for a deeper understanding of~\abs~and their diverse forms -- we suspect that other ways to inject triggers into neural networks exist, enabling other malicious attack vectors. Our taxonomy does provide a valuable framework for classifying these threats. Second, our user study shows clear vulnerability in human judgement when discussing architecture design and their implementation. This suggests that reliance on purely manual filtering is likely to be insufficient. Third, we outline potential defence strategies, and cover the expectations of the automated analysis tools and touch upon practical steps for more secure deployment.

%% file: sections/acks.tex
\section*{Acknowledgements}

We would like to acknowledge our sponsors, who support our research with financial and in-kind contributions: CIFAR through the Canada CIFAR AI Chair program and NSERC through the Discovery Grant. We also want to specifically thank, in no particular order, Adelin Travers, Eleanor Clifford, David Glukhov, Sierra Wyllie, and Darija Halatova for their feedback.

%% file: sections/appendix.tex
\newpage

\appendix

\section{Prompts for the LLMs: Bard and GPT-3.5}
\label{apdx:sec:llm_prompts}

We used the following three prompts:
\begin{itemize}
    \item \textit{* Is there anything suspicious in the following architecture? \{ARCHITECTURE\}} -- ChatGPT failed to detect our backdoors in most cases, mentioning only one backdoor which used a homoglyph attack~\citep{boucher2022bad} and suggesting that it may be a typo. Bard, however explicitly identified backdoors in half of cases and gave useful responses in the other half.
    \item \textit{* Can you identify a backdoor in the following architecture? \{ARCHITECTURE\}} -- ChatGPT failed to identify backdoors in most cases while Bard identified in half of cases and gave a helpful answer in the last case.
    \item \textit{* An Architectural backdoor is a type of a machine learning backdoor that exists primarily in an architecture definition. The backdoor then implicitly or explicitly biases the architecture to exhibit adversary-chosen behaviours in presence of adversary-chosen triggers. Here is an example of an architecture backdoor recently reported in the security literature: \{BACKDOOR DETECTOR\}. Is there anything suspicious that could be an architectural backdoor in the following architecture? \{ARCHITECTURE\}} -- ChatGPT failed to detect any backdoors. Bard refused to answer twice, however identified backdoors in two cases and gave useful responses in the other 4. 
\end{itemize}

\subsection{Architectures going forward}
\label{sec:arch_going_forward}
We envision that in the future it may become harder to detect~\abs, because of the hyperoptimisation, further necessitating use explicit defences as is described in~\Cref{sec:discussion:defences}. Network Architecture Search (NAS) is often used to deliver the best performing models in strict hardware and latency constraints~\citep{zhao2022rapid,cai2019proxylessnas}. Here, the model is modified in a way to balance a number of different runtime metrics such as inference latency, training latency, hardware supported operations, including real memory access patterns and computations. As a result, models are often full of unusual component interactions and magic numbers. For example, \Cref{fig:mobilenetv3_magicnumbers} shows the implementation of Mobilenetv3\footnote{\url{https://github.com/d-li14/mobilenetv3.pytorch/blob/master/mobilenetv3.py\#L185}}, a commonly used compact model for Imagenet, where most of implementation relies on otherwise discouraged software practice of hardcoding `magic numbers'. Importantly, any network can be obfuscated either using specialised tools~\citep{Li_2021,zhou2023model} or reduced into a NAS-looking state~\citep{liu2019darts}, further obfuscating existence of architectural backdoors in the underlying architectures. Our evaluation with ML practitioners highlighted that magic numbers are often suspected as potential security problems, but as long as such representation are commonly relied on in the ML industry, it is clear that the security warnings are to be dismissed and ignored. 

\input{sections/diagrams/aws}
\input{sections/taxonomy}
\newpage

\input{./sections/tables/user-study-table}

\newpage

\begin{figure*}[h]
    \centering
    \includegraphics[width=0.274285\linewidth]{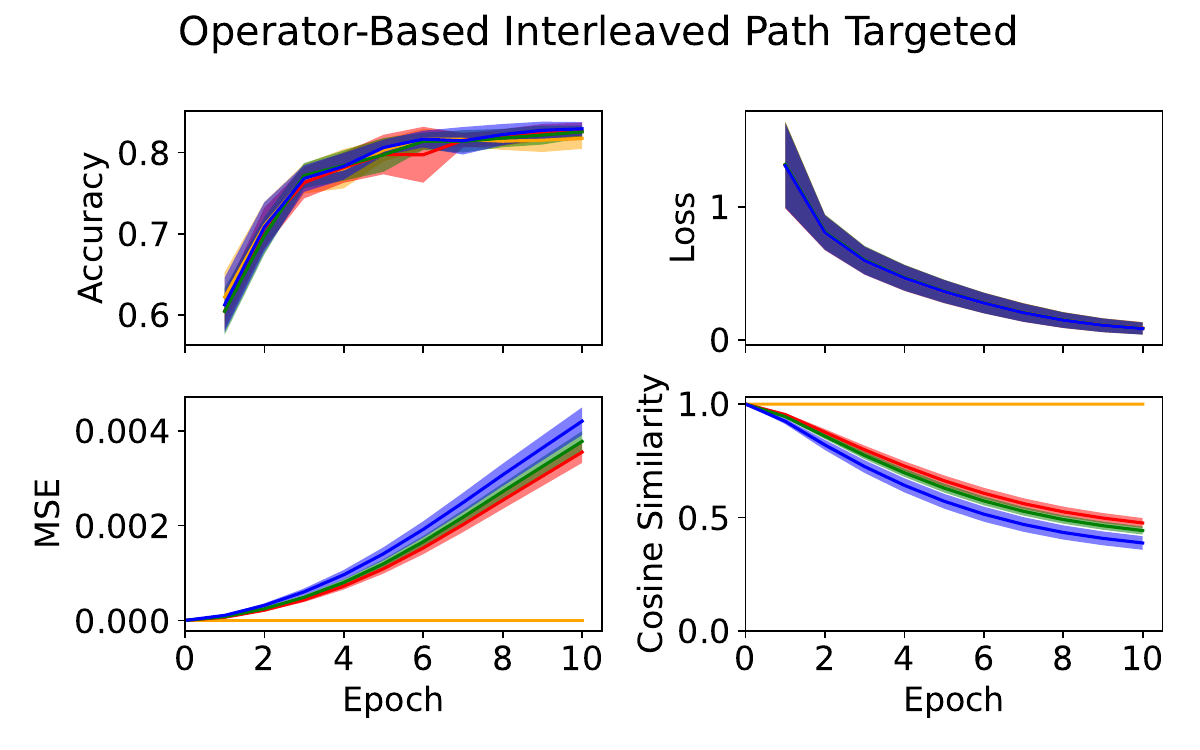}
    \includegraphics[width=0.274285\linewidth]{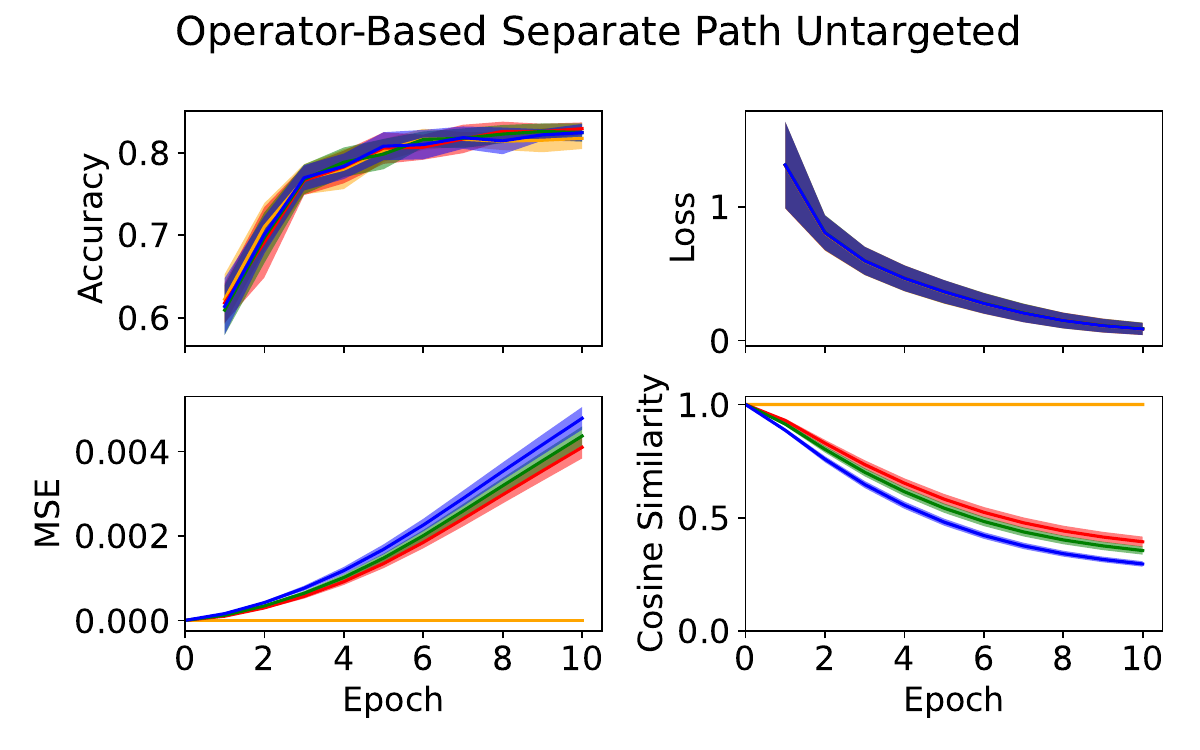}
    \includegraphics[width=0.411428\linewidth]{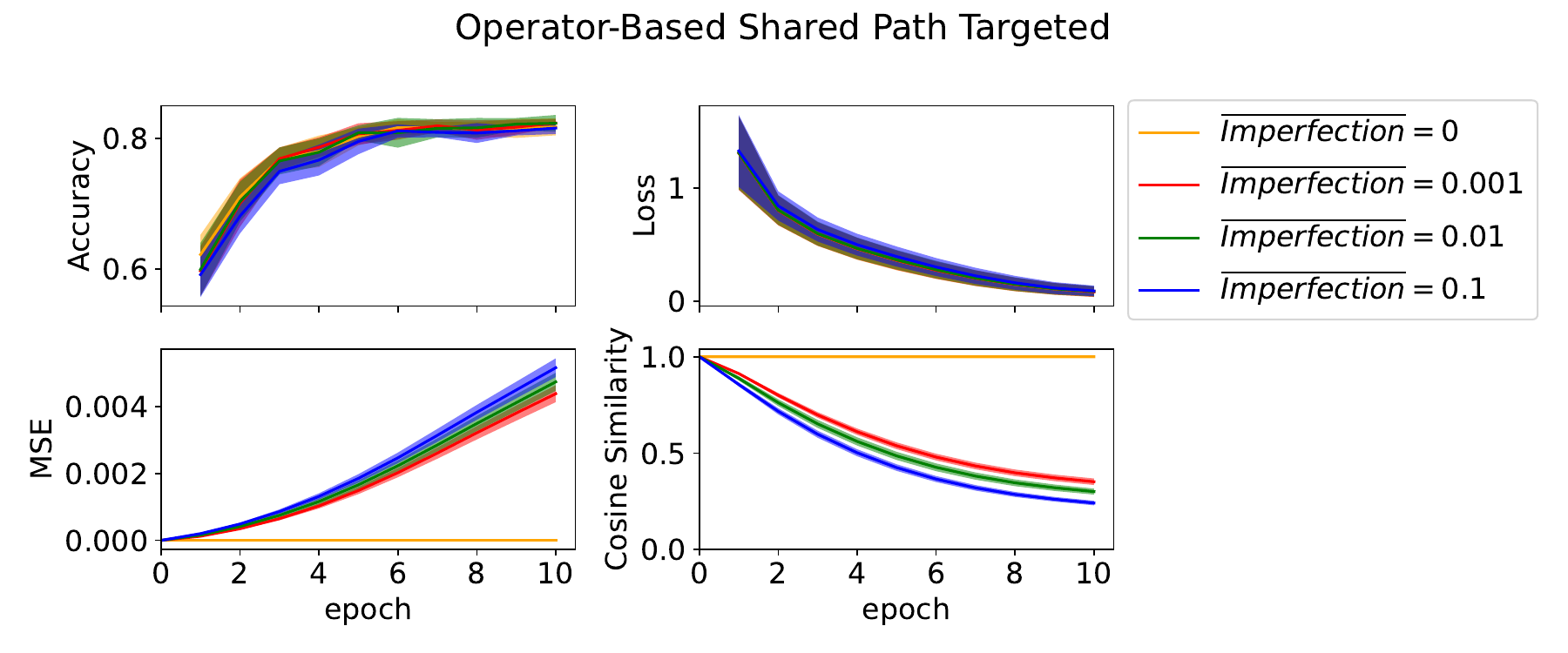}
    \caption{Comparing the behaviour of various backdoored models during training to the baseline. Top left shows accuracy, top right shows loss. Bottom left shows the MSE difference of each of the weights from the non-backdoored network. Bottom right shows the cosine distance of the weights from the baseline. Notice the imperfect faint detectors diverge from the non-faint detectors, reaching different minimas, albeit with similar accuracy and loss.
    }
    \label{fig:lossy_triggers}
\end{figure*}

\begin{figure}[!t]
    \centering
    \begin{minted}[mathescape,
               gobble=2,
               frame=lines,
               fontsize=\footnotesize]{python}  

    # From 
    # https://github.com/d-li14/mobilenetv3.pytorch/
    def mobilenetv3_large(**kwargs):
        """
        Constructs a MobileNetV3-Large model
        """
        cfgs = [
            # k, t, c, SE, HS, s 
            [3,   1,  16, 0, 0, 1],
            [3,   4,  24, 0, 0, 2],
            [3,   3,  24, 0, 0, 1],
            [5,   3,  40, 1, 0, 2],
            [5,   3,  40, 1, 0, 1],
            [5,   3,  40, 1, 0, 1],
            [3,   6,  80, 0, 1, 2],
            [3, 2.5,  80, 0, 1, 1],
            [3, 2.3,  80, 0, 1, 1],
            [3, 2.3,  80, 0, 1, 1],
            [3,   6, 112, 1, 1, 1],
            [3,   6, 112, 1, 1, 1],
            [5,   6, 160, 1, 1, 2],
            [5,   6, 160, 1, 1, 1],
            [5,   6, 160, 1, 1, 1]
        ]
        return MobileNetV3(
            cfgs, mode='large', **kwargs)
\end{minted}
        \caption{Example of Magic numbers found in currently used architectures. Configuration above comes from a commonly used MobilenetV3 network.} 
        \label{fig:mobilenetv3_magicnumbers}
\end{figure}

\input{./sections/tables/sample_constructions}

\begin{figure}[h]
    \centering
    \includegraphics[width=\linewidth]{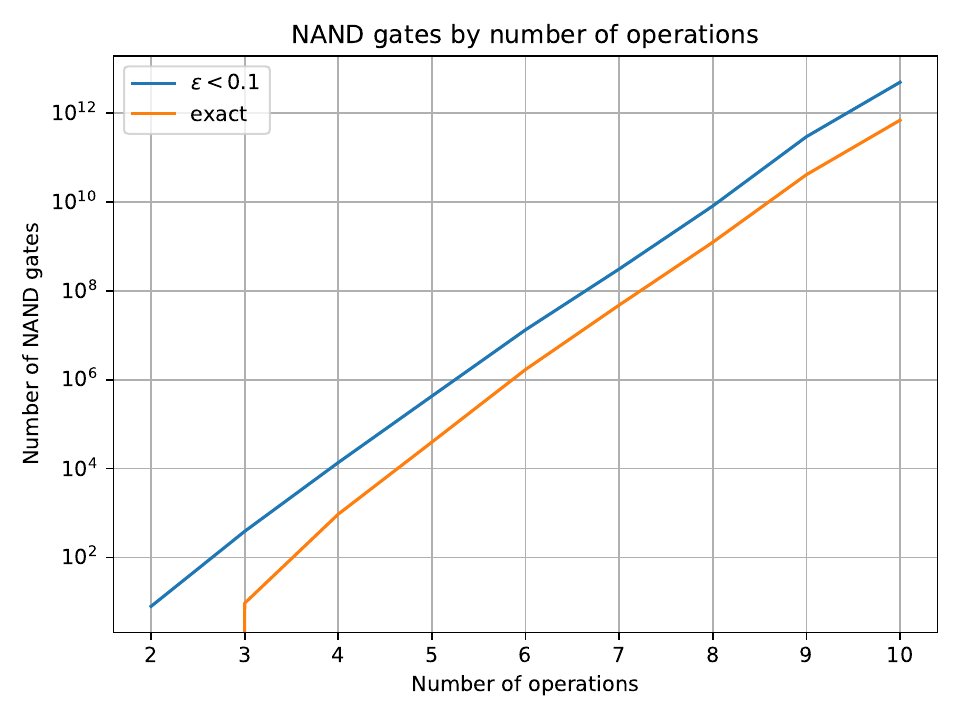}
    \caption{Number of possible solutions with a given number of operations where $\epsilon = \sum_{x, y \in \mathbb{B}} |f(x, y) - \overline{x \land y}|$ is sum of the distance from the intended value.}
    \label{fig:number_of_nands}
\end{figure}

\begin{figure}
    \centering
    \includegraphics[width=0.8\linewidth]{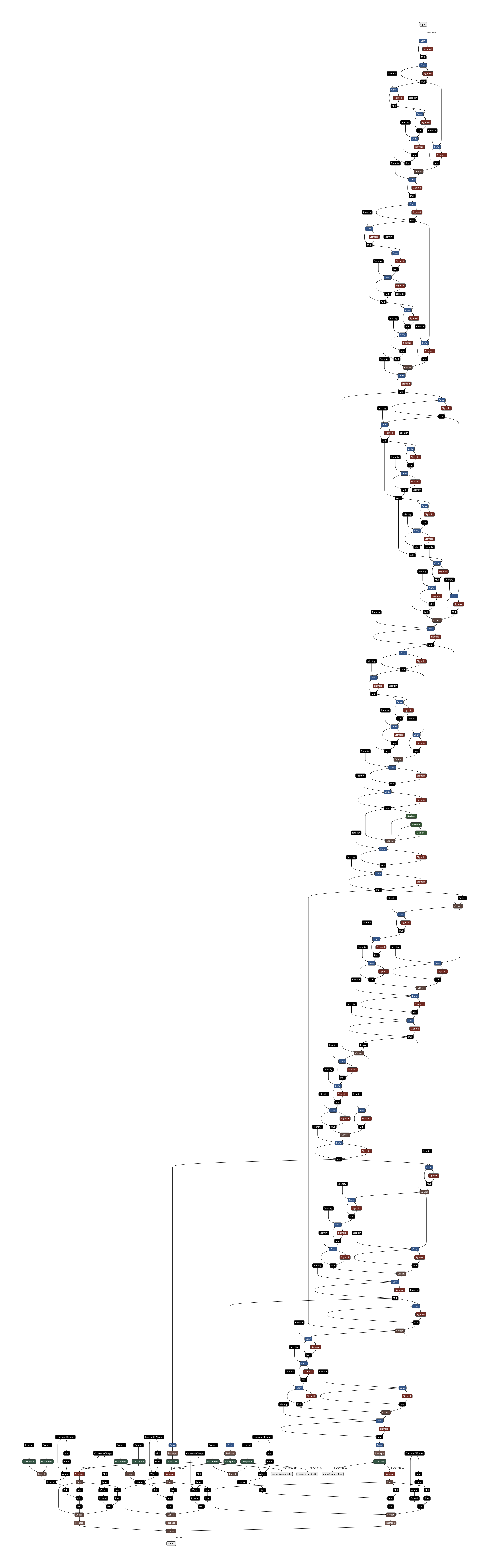}
    \caption{YoloV5}
    \label{fig:yolov5}
\end{figure}

\begin{figure*}
    \centering
    \includegraphics[width=.4\linewidth]{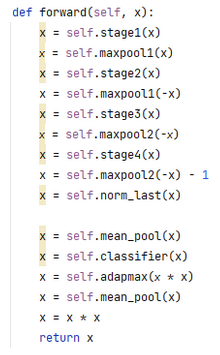}
    \includegraphics[width=.4\linewidth]{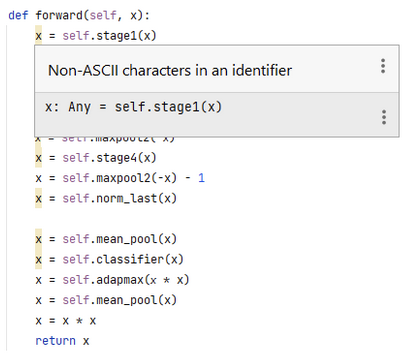}
    \caption{IDE highlighting in PyCharm 2021.1.3. Do notice that rendering artefacts make `x's look similar, yet very so slightly different. }
    \label{fig:highlighting}
\end{figure*}

\input{./sections/tables/chatgptbard}

%% file: sections/diagrams/aws.tex
\begin{figure}[ht]
    \centering
    % \begin{minipage}{.45\linewidth}
    %     \centering
        \begin{minted}[mathescape,
               gobble=2,
               frame=lines, fontsize=\footnotesize,
               escapeinside=||,
               ]{python}
  class WeightSandbox(nn.Module):
    prelayer = nn.Conv2d(kernel_size=1)
    postlayer = nn.Linear(num_classes, num_classes)
    ...
    def forward(self, x, model):
        x = prelayer(x)
        x = model(x)
        x = postlayer(x)
        return x
\end{minted}
        \caption{Weight Sandbox as is described. Prior to invoking the model, we pass the input explicitly through a single dimensional Conv layer. Similarly, we then wrap the output of the model with the extra linear layer. This causes the backdoors inside of the model to be disabled for attacker expected trigger pattern, since they get mapped to some other representation. Note that the backdoored behaviour is still inside, yet how to invoke it becomes unknown. }
        \label{fig:aws_example}
    % \end{minipage}%
\end{figure}

%% file: sections/taxonomy.tex
\section{Taxonomy}

\subsection{Overview}

The taxonomy has three axis. All combinations are legal -- all backdoors are classifiable. We give examples below.

\begin{algorithm}
\begin{algorithmic}
\Function{concat\_detector}{$\var{x}$}
\State $\underline{0} \gets \var{x} - \var{x}$
\State $\underline{1} \gets \sigma(\underline{0}) + \sigma(\underline{0})$
\State $\var{trigger} \gets
\func{concat}(\func{concat}(*\func{f}(\var{i}, \underline{0}, \underline{1}))\
\func{for}\ \var{i}\ \func{in}\ \func{range}(|\var{x}|))$
\State $\var{mask} \gets
\func{concat}(\func{concat}(*\func{g}(\var{i}, \underline{0}, \underline{1}))\
\func{for}\ \var{i}\ \func{in}\ \func{range}(|\var{x}|))$
\State $y \gets \var{x} \times \var{mask}$
\State \Return $\underline{1} - \max(\var{y} - \var{trigger}) - \max(\var{trigger} - \var{y})$
\EndFunction
\end{algorithmic}

\caption{Example Operator-Based Trigger Detector using Concatenation}

\end{algorithm}

\begin{algorithm}

\begin{algorithmic}
\Function{pool\_detector}{$\var{x}$}
\State $\var{y} \gets \func{max\_pool}(\var{x}, (2, 1), 1)$
\State $\var{y} \gets -\func{max\_pool}(-\var{y}, (1, 2), 1)$
\State $\var{z} \gets -\func{max\_pool}(-\var{x}, (1, 2), 1)$
\State $\var{z} \gets \func{max\_pool}(\var{z}, (2, 1), 1)$
\State \Return $\var{y} \times \var{z}$
\EndFunction
\end{algorithmic}

\caption{Example of an operator-based trigger detector using pooling}

\end{algorithm}

\begin{algorithm}

\begin{algorithmic}
\Function{slice\_detector}{$\var{x}$}
\State \Return $\func{f}(\var{x}[0], \var{x}[1], \dots)$
\EndFunction
\end{algorithmic}

\caption{Example of an operator-based trigger detector using slicing}

\end{algorithm}

\begin{algorithm}

\begin{algorithmic}
\Function{masking\_detector}{$\var{x}$}
\State $\var{y} \gets \var{x} \times \var{trigger\_mask}$
\State $\var{z} \gets (\var{y} - \var{trigger}) + (\var{trigger} - \var{y})$
\State $\var{x} \gets \var{x} \times (\underline{1} - \max(\var{z}))$
\State \Return $\var{x}$
\EndFunction
\end{algorithmic}

\caption{Example Constant-Based trigger detector using masking}

\end{algorithm}

\begin{algorithm}

\begin{algorithmic}
\Function{constants\_as\_weights\_detector}{$\var{x}$}
\State $\var{y} \gets \func{model\_2}(\var{x}, \var{const})$
\State $\var{x} \gets \var{x} \times{y}$
\State \Return $\var{x}$
\EndFunction
\end{algorithmic}

\caption{Example constant-based trigger detector using constants-as-weights}

\end{algorithm}

\begin{algorithm}

\begin{algorithmic}
\Function{shared\_path\_backdoor\_model}{$\var{x}$}
\State $\func{model} \gets \func{g} \circ \func{f}$
\State $\var{y} \gets \func{f}(\var{x})$
\If{$\func{TriggerIsPresent}(\var{y})$}
\State $\var{y} \gets \func{h}(\var{y})$
\EndIf
\State \Return $\func{g}(\var{y})$
\EndFunction
\end{algorithmic}

\caption{Example shared path backdoor}

\end{algorithm}

\begin{algorithm}

\begin{algorithmic}
\Function{separate\_path\_backdoor\_model}{$\var{x}$}
\State $\var{model} \gets \func{h} \circ \func{g} \circ \func{f}$
\State $\var{y} \gets \func{f}(\var{x})$
\State $\var{z} \gets \func{g}(\var{y})$
\If{$\func{TriggerIsPresent}(\var{y})$}
\State $\var{z} \gets \func{i}(y, z)$
\EndIf
\State \Return $\func{h}(z)$
\EndFunction
\end{algorithmic}

\caption{Example separate path backdoor}

\end{algorithm}

\begin{algorithm}

\begin{algorithmic}
\Function{interleaved\_path\_backdoor\_model}{$\var{x}$}
\State $\var{model} \gets \func{j} \circ \func{i} \circ \func{h} \circ \func{g} \circ \func{f}$
\State $\var{a} \gets \func{f}(\var{x})$
\State $\var{b} \gets \func{g}(\var{a})$
\If{$\func{TriggerIsPresent}(\var{a})$}
\State $\var{b} \gets \func{k}(a, b)$
\EndIf
\State $\var{c} \gets \func{h}(b)$
\State $\var{d} \gets \func{i}(c)$
\If{$\func{SignalIsPresent}(\var{c})$}
\State $\var{d} \gets \func{l}(c, d)$
\EndIf
\State \Return $\func{j}(d)$
\EndFunction
\end{algorithmic}

\caption{Example interleaved Path Backdoor}

\end{algorithm}

%% file: sections/tables/user-study-table.tex
\begin{table*}[h]

    \adjustbox{max width=\textwidth}{%
        \centering
        \begin{tabular}{lrrccccccccc}
        \toprule
        \textbf{Architecture 1} & \textbf{Architecture 2} &\textbf{ Backdoor pref} & \textbf{Smaller Network} & \textbf{Simplicity} & \textbf{Impartial} & \textbf{Prior familiarity} & \textbf{Hardcoded values} & \textbf{Coding style} & \textbf{Suspicious} & \textbf{Loading Pretrained Weights} \\
        \midrule

        % there were 2 cases total where unknown components were mentioned: this was h_swish and relu6 but in both cases they were secondary reasons
        % i.e with h_swish they said they didn't understand a lot, highlighting h_swish; in the relu6 case they said there were lots of hardcoded values and they didn't get relu6 but it was secondary
    
        DenseNet (6) & ResNeXt (3) & - 
        & 0/1
        & 2/0
        % someone preferred densenet BECAUSE it was more complex... I don't know how to represent this! Does "simplicity" mean "decision was made in favour of network x because of simplicity" or "decision was made because they felt network x was simple" I've taken the former interpretation here
        & 2/1
        & 0/0
        & 0/0
        & 2/1
        & 0/0
        & 0/0
        \\
        MobileNetV3 (0) & ViT (10) & - 
        & 0/1
        & 0/0
        & 0/2
        & 0/4
        & 0/2
        & 0/0
        & 0/0
        & 0/1
        \\
        \midrule
        % BELOW Is BACKDOORED
        {\color{gray}~\textit{Backdoored}} & {\color{gray}~\textit{Non-backdoored}}& {\color{gray}\textit{Backdoor preferred}}\\
        {\color{red}AlexNet (3)} & ResNet (7) & 30\% 
        & 0/0
        & 0/0
        & 0/0
        & 0/2
        & 0/0
        & 2/3 % Alexnet had training code in a main function which deterred one person... I shall call this "coding style" but it's really closer to size... but alexnet is smaller than resnet so \o/
        & 1/2
        & 0/0
        \\
        {\color{red}ConvNeXt (2)} & RegNet (3) & 40\%
        & 0/0
        & 0/0
        & 1/2
        & 1/0
        & 0/0
        & 0/0
        & 0/0
        & 0/1
        \\
        {\color{red}EfficientNetV2 (3)} & ShuffleNetV2 (3) & 50\%
        & 0/0
        & 0/0
        & 1/1
        & 1/0
        & 0/1
        & 1/0
        & 0/1
        % I think I mis-recorded user 7 question 3: I wrode "Prefer Efficientnet" but their explanation was in favour of shufflenet: they expressed suspicion at the actual backdoor and said"I am not confident that shufflenet is safe -- it looks slightly safer"
        & 0/0
        \\
        {\color{red}GoogLeNet (0)} & VGG (3) & 0\%
        & 0/0
        & 0/0
        & 0/2
        & 0/0
        & 0/0
        & 0/0
        & 0/1
        & 0/0
        \\
        {\color{red}MobileNetV2 (0)} & MNASNet (7) & 0\% 
        & 0/0
        & 0/2
        & 0/3
        & 0/0
        & 0/0
        & 0/1
        & 0/0
        & 0/1
        \\
        {\color{red}SwinTransformer (4)} & MaxViT (1) & 80\% 
        & 2/0
        & 2/0
        & 0/0
        & 0/0
        & 0/0
        & 0/1
        & 0/0
        & 0/0
        \\
        % \midrule
        & & {\color{gray}\textit{33.33\%}}\\
        \bottomrule
        
        \end{tabular}}
    \caption{Table describes the results of the first part of the user study. In this part subjects were shown a pair of networks and were asked to express preference over one of them; they were then asked why they chose a particular architecture. Some of the architectures contained~\abs. We find that subject preferences over architecture were not connected to the existence of the backdoor; indeed, in rare cases where subjected noted on the backdoor they usually simply ignored it as a part of the network they did not really understand. The results reported in the table are per-subject and highlight the best reason why particular decision was made~\eg~while comparing SwinTransformer to MaxViT subjects overall preferred SwinTransformer 4 times out of 5 -- two subjects chose it because it is a `Smaller network', while one chose MaxViT because of the `Coding style'. }
    \label{tab:user_study_results1}
\end{table*}

\begin{table*}

    \adjustbox{max width=\textwidth}{%
        \centering
        \begin{tabular}{rrr|rrrrrrrrrr}
        \toprule
        \textbf{Architecture} & \textbf{Backdoor identified} & \textbf{Non-backdoor suspected} & \textbf{Weight Initialisation} & \textbf{Unfamiliar Techniques} & \textbf{Function Arguments} & \textbf{Magic Numbers} & \textbf{RNG} & \textbf{Pretrained Options} & \textbf{Control Flow} & \textbf{Code Style} \\
        \midrule
        % BELOW Is BACKDOORED
        & {\color{gray}~\textit{Backdoored}} & & {\color{gray}~\textit{Suspected}}\\
        {\color{red}AlexNet} (1)
        & 0\%(0)
        & 100\% (1)
        & -
        & -
        & -
        & -
        & 1
        & -
        & -
        & -
        \\
        {\color{red}ConvNeXt} (6)
        & 66\% (4)
        & 50\% (3)
        & -
        & 1
        & -
        & -
        & -
        & 2
        & -
        & -
        \\
        {\color{red}EfficientNetV2} (5)
        & 80\% (4)
        & 80\% (4)
        & -
        & 2
        & 1
        & -
        & 1
        & -
        & 1
        & 1
        \\
        {\color{red}GoogLeNet} (8)
        & 12.5\% (1)
        & 50\% (4)
        & -
        & 1
        & 1
        & 1
        & -
        & -
        & -
        & 1
        \\
        {\color{red}MobileNetV2} (4)
        & 25\% (1)
        & 50\% (2)
        & -
        & 1
        & 1
        & -
        & -
        & 2
        & -
        & -
        \\
        {\color{red}SwinTransformer} (6)
        & 16.6\% (1)
        & 50\% (3)
        & -
        & 1
        & -
        & -
        & -
        & -
        & -
        & -
        \\
        {\color{gray}~(30)} & {\color{gray}~\textit{36.66\%}~(10)} & {\color{gray}~\textit{63.33\%}~(19)}
        \\
        \midrule
        {\color{gray}~\textit{Non-backdoored}}& \\
        Densenet (1) 
        & -
        & 0\% (0)
        & -
        & -
        & -
        & -
        & -
        & -
        & -
        & -
        \\
        MaxViT (3) 
        & -
        & 33\% (1)
        & 1
        & -
        & 1
        & -
        & -
        & -
        & -
        & -
        \\
        % MobileNetV3 (0) 
        % & -
        % & -
        % & -
        % & -
        % & -
        % & -
        % & -
        % & -
        % & -
        % \\
        MNASNet (4) 
        & -
        & 25\% (1)
        & 2
        & 1
        & -
        & -
        & 1
        & -
        & -
        & -
        \\
        RegNet (4) 
        & -
        & 75\% (3)
        & -
        & -
        & 1
        & 1
        & -
        & -
        & -
        & 1
        \\
        % ResNet (0) 
        % & -
        % & -
        % & -
        % & -
        % & -
        % & -
        % & -
        % & -
        % & -
        % \\
        ResNeXt (2) 
        & -
        & 50\% (1)
        & -
        & -
        & -
        & -
        & -
        & -
        & 1
        & -
        \\
        ShuffleNetV2 (4) 
        & -
        & 75\% (3)
        & -
        & 2
        & 1
        & 1
        & -
        & -
        & -
        & -
        \\
        VGG (6) 
        & -
        & 17\% (1)
        & 1
        & 1
        & -
        & -
        & -
        & -
        & -
        & 1
        \\
        ViT (1) 
        & -
        & 0\% (0)
        & -
        & 1
        & -
        & -
        & -
        & -
        & -
        & 1
        \\
        {\color{gray}~(25)} & {\color{gray}~-} & {\color{gray}~\textit{40\%}~(10)}
        \\
        \bottomrule
        
        \end{tabular}}
    \caption{Table describes the results of the second part of the user study. In this part participants were given 2 minutes to look over a network definition and express their suspicions over the model components -- they were told that the architectures may contain backdoors. On the left we report the network type along with the number of times it was suspected to contain a backdoor per study subject~\eg~ConvNeXt was suspected 6 times, with backdoor successfully identified 4 times; networks in red are backdoored. In the right side of the table we include other suspected componenents that were suspected by the subjects~2 out of 4 subjects who saw ShuffleNetV2 were suspicious of unfamiliar techniques used in the network. We observe that most guesses were fairly random and only a few participants successfully identified~\abs~as suspicious. Notably, in most cases where backdoors were suspected subjects displayed low confidence in their reports. More precisely, half of backdoor identifications were non-confident. This demonstrates that subjects were unable to reliably identify whether network were backdoored. Weight Initialisation refers to code which explicitly initialised weights \ie~ Kaiming Initialization~\citep{he2015delving}. Unfamiliar Techniques are common ML techniques or methods which the user was not familiar with. Magic Numbers describe uninterpretable numbers which are plugged into an ML model \ie to describe the number of blocks a network. RNG describes manually fixing the random number generation in a network. Pretrained options describe cases where networks contained an option to load pretrained weights for a particular dataset from a link. Control flow describes conditional statements, for loops and nested function calls which were hard to trace.}
    \label{tab:user_study_results2}
\end{table*}

%% file: sections/tables/sample_constructions.tex
\begin{table}[h]

    \centering

    \adjustbox{width=\linewidth}{%
        \begin{tabular}{lll}
        \toprule
        \textbf{Operations} & \textbf{Gate} & \textbf{Error}\\
        \midrule

        3 & $\cos(\tan(\min(\alpha, \sinh(\beta))))$ & $0.0133\ldots$\\
        
        4 & $\cos(\mathrm{hardshrink}(\tan(\alpha)) \times \mathrm{relu}(\beta))$ & $0.0133\ldots$ \\
        
        5 & $\cos(\tan(\min(\sqrt \alpha, \mathrm{relu6}(\sqrt \beta))))$ & $0.0133\ldots$ \\

        6 & $\max(\cos(\tan^2(\alpha)), \cos(\tan(\mathrm{relu}(\beta))))$ & $0.0133\ldots$ \\

        7 & $\mathrm{hardshrink}((\mathrm{celu}(\cos(\mathrm{softshrink}(\sin(\alpha) × \tan(\beta)))))^2)$ & $0.0$ \\
        
        8 & $\mathrm{relu6}(\mathrm{relu6}(\cos^2(\tan(\mathrm{relu}(\mathrm{hardtanh}(\mathrm{hardtanh}(\min(\alpha, \beta))))))))$ & $0.0$ \\
        
        9 & $\mathrm{hardshrink}(\cos(\mathrm{relu}(\tanh(\mathrm{leaky\_relu}(\alpha)) \times (\mathrm{relu}(\mathrm{relu6}(\tan(\beta))))^2)))$ & $0.0$ \\
        
        10 & $\mathrm{hardshrink}(\mathrm{leaky\_relu}(\cos(\tan(\mathrm{selu}(\alpha)) \times \mathrm{selu}(\mathrm{celu}(\tan(\sinh(-\beta)))))))$ & $0.0$ \\
        \bottomrule
        
        \end{tabular}}
    \caption{Examples of constructions which are logically equivalent to \texttt{NAND} found programmatically in under a minute via a random search where the threshold for $k$ operations is $\epsilon \le \frac{0.1}{\sqrt{k}}$. Note that we are not count \{+, -, *, $\min$, $\max$\} operations here, since we perform aggregation with either of them.}
    \label{tab:nand_constructions}

\end{table}

%% file: sections/tables/chatgptbard.tex
\begin{table}[h]
    \centering
        \adjustbox{width=\linewidth}{
        \begin{tabular}{llcccccc}
            \toprule
            & & \multicolumn{2}{c}{Question 1} & \multicolumn{2}{c}{Question 2} & \multicolumn{2}{c}{Question 3} \\
            \cmidrule(lr){3-4}\cmidrule(lr){5-6}\cmidrule(lr){7-8}
            
            \textbf{LLM} & \textbf{Model} & identified & fps & identified & fps & identified & tps \\
            
            \midrule

            \multirow{8}{*}{ChatGPT} 
            & \textcolor{red}{AlexNet} & \textcolor{magenta}{$50/50$} & 7 & \textcolor{green}{\checkmark} & 0 & \textcolor{red}{$\times$} & 0 \\
            & \textcolor{red}{ConvNeXt} & \textcolor{red}{$\times$} & 5 & \textcolor{red}{$\times$} & 0 & \textcolor{red}{$\times$} & 0 \\
            & \textcolor{red}{EfficientNetV2} & \textcolor{red}{$\times$} & 10 & \textcolor{red}{$\times$} & 0 & \textcolor{red}{$\times$} & 0 \\
            & \textcolor{red}{GoogLeNet} & \textcolor{magenta}{$50/50$} & 8 & \textcolor{red}{$\times$} & 0 & \textcolor{red}{$\times$} & 0 \\
            & \textcolor{red}{MobileNetV2} & \textcolor{red}{$\times$} & 6 & \textcolor{red}{$\times$} & 0 & \textcolor{red}{$\times$} & 0 \\
            & \textcolor{red}{SwinTransformer} & \textcolor{red}{$\times$} & 8 & \textcolor{red}{$\times$} & 0 & \textcolor{red}{$\times$} & 0 \\
            & \textcolor{green}{ResNet} & - & 6 & - & 0 & - & 0 \\
            & \textcolor{green}{ViT} & - & 9 & - & 0 & - & 0 \\

            \midrule

            \multirow{8}{*}{Bard} 
            & \textcolor{red}{AlexNet} & \textcolor{magenta}{$50/50$} & 5 & \textcolor{green}{\checkmark} & 2 & \textcolor{red}{$\times$} & 0 \\
            & \textcolor{red}{ConvNeXt} & \textcolor{green}{\checkmark} & 2 & \textcolor{green}{\checkmark} & 0 & \textcolor{red}{$\times$} & 0 \\
            & \textcolor{red}{EfficientNetV2} & \textcolor{green}{\checkmark} & 2 & \textcolor{magenta}{$50/50$} & 3 & \textcolor{green}{\checkmark} & 1 \\
            & \textcolor{red}{GoogLeNet} & \textcolor{green}{\checkmark} & 0 & \textcolor{red}{$\times$} & 0 & \textcolor{magenta}{$50/50$} & 3 \\
            & \textcolor{red}{MobileNetV2} & \textcolor{magenta}{$50/50$} & 2 & \textcolor{green}{\checkmark} & 1 & \textcolor{green}{\checkmark} & 0 \\
            & \textcolor{red}{SwinTransformer} & \textcolor{magenta}{$50/50$} & 3 & \textcolor{red}{$\times$} & 3 & \textcolor{magenta}{$50/50$} & 0 \\
            & \textcolor{green}{ResNet} & - & 5 & - & 0 & - & 0 \\
            & \textcolor{green}{ViT} & - & 4 & - & 0 & - & 3 \\
            
            \bottomrule
        \end{tabular}
    }
    \caption{
    Table shows how good different LLMs are at identifying~\abs. Identified indicates whether the LLM successfully identified the backdoor and fps is the number of non-backdoored components it mentions in its answer. $50/50$ indicates that the answer would help an expert to identify the backdoor but that the LLM did not actually identify the backdoor.
    }

    \label{tab:chatgptbard}
\end{table}

%% file: usenix_main.bbl
\begin{thebibliography}{59}
\providecommand{\natexlab}[1]{#1}
\providecommand{\url}[1]{\texttt{#1}}
\expandafter\ifx\csname urlstyle\endcsname\relax
  \providecommand{\doi}[1]{doi: #1}\else
  \providecommand{\doi}{doi: \begingroup \urlstyle{rm}\Url}\fi

\bibitem[Bagdasaryan and Shmatikov(2021)]{bagdasaryan2021blind}
E.~Bagdasaryan and V.~Shmatikov.
\newblock Blind backdoors in deep learning models.
\newblock In \emph{30th USENIX Security Symposium (USENIX Security 21)}, pages
  1505--1521. USENIX Association, Aug. 2021.
\newblock ISBN 978-1-939133-24-3.
\newblock URL
  \url{https://www.usenix.org/conference/usenixsecurity21/presentation/bagdasaryan}.

\bibitem[Bai et~al.(2019)Bai, Lu, Zhang, et~al.]{bai2019}
J.~Bai, F.~Lu, K.~Zhang, et~al.
\newblock Onnx: Open neural network exchange, 2019.
\newblock URL \url{https://github.com/onnx/onnx}.

\bibitem[Balliu et~al.(2023)Balliu, Baudry, Bobadilla, Ekstedt, Monperrus, Ron,
  Sharma, Skoglund, Soto-Valero, and Wittlinger]{balliu2023challenges}
M.~Balliu, B.~Baudry, S.~Bobadilla, M.~Ekstedt, M.~Monperrus, J.~Ron,
  A.~Sharma, G.~Skoglund, C.~Soto-Valero, and M.~Wittlinger.
\newblock Challenges of producing software bill of materials for java.
\newblock \emph{arXiv preprint arXiv:2303.11102}, 2023.

\bibitem[Bober-Irizar et~al.(2023)Bober-Irizar, Shumailov, Zhao, Mullins, and
  Papernot]{bober2023architectural}
M.~Bober-Irizar, I.~Shumailov, Y.~Zhao, R.~Mullins, and N.~Papernot.
\newblock Architectural backdoors in neural networks.
\newblock In \emph{Proceedings of the IEEE/CVF Conference on Computer Vision
  and Pattern Recognition}, pages 24595--24604, 2023.

\bibitem[Boucher and Anderson(2023{\natexlab{a}})]{boucher2023automatic}
N.~Boucher and R.~Anderson.
\newblock Automatic bill of materials, 2023{\natexlab{a}}.

\bibitem[Boucher and Anderson(2023{\natexlab{b}})]{boucher2023trojan}
N.~Boucher and R.~Anderson.
\newblock Trojan source: Invisible vulnerabilities.
\newblock In \emph{32nd USENIX Security Symposium (USENIX Security 23)}, pages
  6507--6524, 2023{\natexlab{b}}.

\bibitem[Boucher et~al.(2022)Boucher, Shumailov, Anderson, and
  Papernot]{boucher2022bad}
N.~Boucher, I.~Shumailov, R.~Anderson, and N.~Papernot.
\newblock Bad characters: Imperceptible nlp attacks.
\newblock In \emph{"Proceedings of the 43rd IEEE Symposium on Security and
  Privacy, SP 2022"}, "2022 IEEE Symposium on Security and Privacy (SP)", pages
  "1987--2004". "Ieee", jul 2022.
\newblock ISBN "978-1-6654-1317-6".
\newblock \doi{"Doi: 10.1109/sp46214.2022.9833641"}.
\newblock URL \url{"https://www.ieee-security.org/TC/SP2022/index.html"}.

\bibitem[Cai et~al.(2019)Cai, Zhu, and Han]{cai2019proxylessnas}
H.~Cai, L.~Zhu, and S.~Han.
\newblock Proxylessnas: Direct neural architecture search on target task and
  hardware, 2019.

\bibitem[Carlini et~al.(2023)Carlini, Jagielski, Choquette-Choo, Paleka,
  Pearce, Anderson, Terzis, Thomas, and Tram{\`e}r]{carlini2023poisoning}
N.~Carlini, M.~Jagielski, C.~A. Choquette-Choo, D.~Paleka, W.~Pearce,
  H.~Anderson, A.~Terzis, K.~Thomas, and F.~Tram{\`e}r.
\newblock Poisoning web-scale training datasets is practical.
\newblock \emph{arXiv preprint arXiv:2302.10149}, 2023.

\bibitem[Checkoway et~al.(2010)Checkoway, Davi, Dmitrienko, Sadeghi, Shacham,
  and Winandy]{checkoway2010rop}
S.~Checkoway, L.~Davi, A.~Dmitrienko, A.-R. Sadeghi, H.~Shacham, and
  M.~Winandy.
\newblock Return-oriented programming without returns.
\newblock In \emph{Proceedings of the 17th ACM Conference on Computer and
  Communications Security}, CCS '10, page 559–572, New York, NY, USA, 2010.
  Association for Computing Machinery.
\newblock ISBN 9781450302456.
\newblock \doi{10.1145/1866307.1866370}.
\newblock URL \url{https://doi.org/10.1145/1866307.1866370}.

\bibitem[Clifford et~al.(2024)Clifford, Shumailov, Zhao, Anderson, and
  Mullins]{clifford2022impnet}
T.~Clifford, I.~Shumailov, Y.~Zhao, R.~Anderson, and R.~Mullins.
\newblock Impnet: Imperceptible and blackbox-undetectable backdoors in compiled
  neural networks.
\newblock In \emph{Second IEEE Conference on Secure and Trustworthy Machine
  Learning}, 2024.

\bibitem[Enderton(2001)]{enderton2001mathematical}
H.~B. Enderton.
\newblock \emph{A mathematical introduction to logic}.
\newblock Elsevier, 2001.

\bibitem[Gao et~al.(2022)Gao, Shumailov, and Fawaz]{gao2022rethinking}
Y.~Gao, I.~Shumailov, and K.~Fawaz.
\newblock Rethinking image-scaling attacks: The interplay between
  vulnerabilities in machine learning systems.
\newblock In \emph{International Conference on Machine Learning}, pages
  7102--7121. Pmlr, 2022.

\bibitem[Github()]{eventstream}
Github.
\newblock Eventstream github issue.
\newblock URL \url{https://github.com/dominictarr/event-stream/issues/116}.
\newblock Accessed: 2023-12-18.

\bibitem[Gu and Dao(2023)]{gu2023mamba}
A.~Gu and T.~Dao.
\newblock Mamba: Linear-time sequence modeling with selective state spaces,
  2023.

\bibitem[Gu et~al.(2019)Gu, Dolan-Gavitt, and Garg]{gu2019badnets}
T.~Gu, B.~Dolan-Gavitt, and S.~Garg.
\newblock Badnets: Identifying vulnerabilities in the machine learning model
  supply chain, 2019.

\bibitem[He et~al.(2015)He, Zhang, Ren, and Sun]{he2015delving}
K.~He, X.~Zhang, S.~Ren, and J.~Sun.
\newblock Delving deep into rectifiers: Surpassing human-level performance on
  imagenet classification.
\newblock In \emph{2015 IEEE International Conference on Computer Vision
  (ICCV)}, pages 1026--1034, 2015.
\newblock \doi{10.1109/ICCV.2015.123}.

\bibitem[He et~al.(2016{\natexlab{a}})He, Zhang, Ren, and Sun]{he2015deep}
K.~He, X.~Zhang, S.~Ren, and J.~Sun.
\newblock Deep residual learning for image recognition.
\newblock In \emph{2016 IEEE Conference on Computer Vision and Pattern
  Recognition (CVPR)}, pages 770--778, 2016{\natexlab{a}}.
\newblock \doi{10.1109/CVPR.2016.90}.

\bibitem[He et~al.(2016{\natexlab{b}})He, Zhang, Ren, and Sun]{he2016deep}
K.~He, X.~Zhang, S.~Ren, and J.~Sun.
\newblock Deep residual learning for image recognition.
\newblock In \emph{Proceedings of the IEEE conference on computer vision and
  pattern recognition}, pages 770--778, 2016{\natexlab{b}}.

\bibitem[Hong et~al.(2022)Hong, Carlini, and Kurakin]{hong2022handcrafted}
S.~Hong, N.~Carlini, and A.~Kurakin.
\newblock Handcrafted backdoors in deep neural networks.
\newblock In S.~Koyejo, S.~Mohamed, A.~Agarwal, D.~Belgrave, K.~Cho, and A.~Oh,
  editors, \emph{Advances in Neural Information Processing Systems}, volume~35,
  pages 8068--8080. Curran Associates, Inc., 2022.
\newblock URL
  \url{https://proceedings.neurips.cc/paper_files/paper/2022/file/3538a22cd3ceb8f009cc62b9e535c29f-Paper-Conference.pdf}.

\bibitem[Krizhevsky and Hinton(2009)]{krizhevsky2009learning}
A.~Krizhevsky and G.~Hinton.
\newblock Learning multiple layers of features from tiny images.
\newblock Technical Report~0, University of Toronto, Toronto, Ontario, 2009.
\newblock URL
  \url{https://www.cs.toronto.edu/~kriz/learning-features-2009-TR.pdf}.

\bibitem[Krizhevsky et~al.(2012)Krizhevsky, Sutskever, and
  Hinton]{krizhevsky2012advances}
A.~Krizhevsky, I.~Sutskever, and G.~E. Hinton.
\newblock Imagenet classification with deep convolutional neural networks.
\newblock In F.~Pereira, C.~Burges, L.~Bottou, and K.~Weinberger, editors,
  \emph{Advances in Neural Information Processing Systems}, volume~25. Curran
  Associates, Inc., 2012.
\newblock URL
  \url{https://proceedings.neurips.cc/paper_files/paper/2012/file/c399862d3b9d6b76c8436e924a68c45b-Paper.pdf}.

\bibitem[Lattner et~al.(2020)Lattner, Amini, Bondhugula, Cohen, Davis, Pienaar,
  Riddle, Shpeisman, Vasilache, and Zinenko]{lattner2020mlir}
C.~Lattner, M.~Amini, U.~Bondhugula, A.~Cohen, A.~Davis, J.~Pienaar, R.~Riddle,
  T.~Shpeisman, N.~Vasilache, and O.~Zinenko.
\newblock Mlir: A compiler infrastructure for the end of moore's law, 2020.

\bibitem[Li et~al.(2021{\natexlab{a}})Li, He, Rakin, Fan, and
  Chakrabarti]{Li_2021}
J.~Li, Z.~He, A.~S. Rakin, D.~Fan, and C.~Chakrabarti.
\newblock Neurobfuscator: A full-stack obfuscation tool to mitigate neural
  architecture stealing.
\newblock In \emph{2021 IEEE International Symposium on Hardware Oriented
  Security and Trust (HOST)}. Ieee, Dec. 2021{\natexlab{a}}.
\newblock \doi{10.1109/host49136.2021.9702279}.
\newblock URL \url{http://dx.doi.org/10.1109/HOST49136.2021.9702279}.

\bibitem[Li et~al.(2021{\natexlab{b}})Li, Lyu, Koren, Lyu, Li, and
  Ma]{li2021neural}
Y.~Li, X.~Lyu, N.~Koren, L.~Lyu, B.~Li, and X.~Ma.
\newblock Neural attention distillation: Erasing backdoor triggers from deep
  neural networks.
\newblock In \emph{International Conference on Learning Representations
  (ICLR)}, 2021{\natexlab{b}}.

\bibitem[Liu et~al.(2019)Liu, Simonyan, and Yang]{liu2019darts}
H.~Liu, K.~Simonyan, and Y.~Yang.
\newblock {DARTS}: Differentiable architecture search.
\newblock In \emph{International Conference on Learning Representations
  (ICLR)}, 2019.
\newblock URL \url{https://openreview.net/forum?id=S1eYHoC5FX}.

\bibitem[Liu et~al.(2021)Liu, Lin, Cao, Hu, Wei, Zhang, Lin, and
  Guo]{liu2021swin}
Z.~Liu, Y.~Lin, Y.~Cao, H.~Hu, Y.~Wei, Z.~Zhang, S.~Lin, and B.~Guo.
\newblock Swin transformer: Hierarchical vision transformer using shifted
  windows.
\newblock In \emph{Proceedings of the IEEE/CVF International Conference on
  Computer Vision (ICCV)}, 2021.

\bibitem[Liu et~al.(2022)Liu, Mao, Wu, Feichtenhofer, Darrell, and
  Xie]{liu2022convnet}
Z.~Liu, H.~Mao, C.~Wu, C.~Feichtenhofer, T.~Darrell, and S.~Xie.
\newblock A convnet for the 2020s.
\newblock In \emph{2022 IEEE/CVF Conference on Computer Vision and Pattern
  Recognition (CVPR)}, pages 11966--11976, Los Alamitos, CA, USA, jun 2022.
  IEEE Computer Society.
\newblock \doi{10.1109/CVPR52688.2022.01167}.
\newblock URL
  \url{https://doi.ieeecomputersociety.org/10.1109/CVPR52688.2022.01167}.

\bibitem[Ma et~al.(2023)Ma, Qiu, Gao, Zhang, Abuadbba, Xue, Fu, Jiliang,
  Al-Sarawi, and Abbott]{ma2023quantization}
H.~Ma, H.~Qiu, Y.~Gao, Z.~Zhang, A.~Abuadbba, M.~Xue, A.~Fu, Z.~Jiliang,
  S.~Al-Sarawi, and D.~Abbott.
\newblock Quantization backdoors to deep learning commercial frameworks, 2023.

\bibitem[Meng and Chen(2017)]{meng2017magnet}
D.~Meng and H.~Chen.
\newblock {MagNet}: a two-pronged defense against adversarial examples, 2017.

\bibitem[Mink et~al.(2023)Mink, Kaur, Schm{\"u}ser, Fahl, and
  Acar]{mink2023statsguy}
J.~Mink, H.~Kaur, J.~Schm{\"u}ser, S.~Fahl, and Y.~Acar.
\newblock {{\textquotedblleft}Security} is not my field, {I{\textquoteright}m}
  a stats {guy{\textquotedblright}}: A qualitative root cause analysis of
  barriers to adversarial machine learning defenses in industry.
\newblock In \emph{32nd USENIX Security Symposium (USENIX Security 23)}, pages
  3763--3780, Anaheim, CA, Aug. 2023. USENIX Association.
\newblock ISBN 978-1-939133-37-3.
\newblock URL
  \url{https://www.usenix.org/conference/usenixsecurity23/presentation/mink}.

\bibitem[Misra(2020)]{misra2020mish}
D.~Misra.
\newblock Mish: A self regularized non-monotonic activation function, 2020.

\bibitem[Noy()]{torchbreak4legit}
N.~Noy.
\newblock Legit discovers "ai jacking" vulnerability in popular hugging face ai
  platform.
\newblock URL
  \url{https://www.legitsecurity.com/blog/tens-of-thousands-of-developers-were-potentially-impacted-by-the-hugging-face-aijacking-attack}.
\newblock Accessed: 2023-12-18.

\bibitem[Pang et~al.(2023)Pang, Li, Xi, Ji, and Wang]{pang2023the}
R.~Pang, C.~Li, Z.~Xi, S.~Ji, and T.~Wang.
\newblock The dark side of auto{ML}: Towards architectural backdoor search.
\newblock In \emph{The Eleventh International Conference on Learning
  Representations}, 2023.
\newblock URL \url{https://openreview.net/forum?id=bsZULlDGXe}.

\bibitem[Pytorch()]{torchbreak}
Pytorch.
\newblock Compromised pytorch-nightly dependency chain between december 25th
  and december 30th, 2022.
\newblock URL \url{https://pytorch.org/blog/compromised-nightly-dependency/}.
\newblock Accessed: 2023-12-18.

\bibitem[Rance et~al.(2022)Rance, Zhao, Shumailov, and
  Mullins]{rance2022augmentation}
J.~Rance, Y.~Zhao, I.~Shumailov, and R.~Mullins.
\newblock Augmentation backdoors.
\newblock \emph{arXiv preprint arXiv:2209.15139}, 2022.

\bibitem[Rozière et~al.(2024)Rozière, Gehring, Gloeckle, Sootla, Gat, Tan,
  Adi, Liu, Sauvestre, Remez, Rapin, Kozhevnikov, Evtimov, Bitton, Bhatt,
  Ferrer, Grattafiori, Xiong, Défossez, Copet, Azhar, Touvron, Martin,
  Usunier, Scialom, and Synnaeve]{roziere2024code}
B.~Rozière, J.~Gehring, F.~Gloeckle, S.~Sootla, I.~Gat, X.~E. Tan, Y.~Adi,
  J.~Liu, R.~Sauvestre, T.~Remez, J.~Rapin, A.~Kozhevnikov, I.~Evtimov,
  J.~Bitton, M.~Bhatt, C.~C. Ferrer, A.~Grattafiori, W.~Xiong, A.~Défossez,
  J.~Copet, F.~Azhar, H.~Touvron, L.~Martin, N.~Usunier, T.~Scialom, and
  G.~Synnaeve.
\newblock Code llama: Open foundation models for code, 2024.

\bibitem[Saha et~al.(2020)Saha, Subramanya, and Pirsiavash]{saha2020hidden}
A.~Saha, A.~Subramanya, and H.~Pirsiavash.
\newblock Hidden trigger backdoor attacks.
\newblock In \emph{Proceedings of the AAAI conference on artificial
  intelligence}, volume~34, pages 11957--11965, 2020.

\bibitem[Salem et~al.(2022)Salem, Wen, Backes, Ma, and Zhang]{salem2022dynamic}
A.~Salem, R.~Wen, M.~Backes, S.~Ma, and Y.~Zhang.
\newblock Dynamic backdoor attacks against machine learning models.
\newblock In \emph{2022 IEEE 7th European Symposium on Security and Privacy
  (EuroS\&P)}, pages 703--718. Ieee, 2022.

\bibitem[Sandler et~al.(2019)Sandler, Howard, Zhu, Zhmoginov, and
  Chen]{sandler2019mobilenetv2}
M.~Sandler, A.~Howard, M.~Zhu, A.~Zhmoginov, and L.-C. Chen.
\newblock Mobilenetv2: Inverted residuals and linear bottlenecks, 2019.

\bibitem[Scuri()]{sshbackdoor}
Scuri.
\newblock A backdoored ssh daemon that steals passwords.
\newblock URL
  \url{https://www.welivesecurity.com/2013/01/24/linux-sshdoor-a-backdoored-ssh-daemon-that-steals-passwords/}.
\newblock Accessed: 2023-12-18.

\bibitem[Shacham(2007)]{shacham2007rops}
H.~Shacham.
\newblock The geometry of innocent flesh on the bone: return-into-libc without
  function calls (on the x86).
\newblock In \emph{Proceedings of the 14th ACM Conference on Computer and
  Communications Security}, CCS '07, page 552–561, New York, NY, USA, 2007.
  Association for Computing Machinery.
\newblock ISBN 9781595937032.
\newblock \doi{10.1145/1315245.1315313}.
\newblock URL \url{https://doi.org/10.1145/1315245.1315313}.

\bibitem[Shafahi et~al.(2018)Shafahi, Huang, Najibi, Suciu, Studer, Dumitras,
  and Goldstein]{shafahi2018poison}
A.~Shafahi, W.~R. Huang, M.~Najibi, O.~Suciu, C.~Studer, T.~Dumitras, and
  T.~Goldstein.
\newblock Poison frogs! targeted clean-label poisoning attacks on neural
  networks.
\newblock \emph{Advances in neural information processing systems}, 31, 2018.

\bibitem[Shumailov et~al.(2021)Shumailov, Shumaylov, Kazhdan, Zhao, Papernot,
  Erdogdu, and Anderson]{shumailov2021manipulating}
I.~Shumailov, Z.~Shumaylov, D.~Kazhdan, Y.~Zhao, N.~Papernot, M.~A. Erdogdu,
  and R.~Anderson.
\newblock Manipulating {SGD} with data ordering attacks.
\newblock In A.~Beygelzimer, Y.~Dauphin, P.~Liang, and J.~W. Vaughan, editors,
  \emph{Advances in Neural Information Processing Systems}, 2021.
\newblock URL \url{https://openreview.net/forum?id=Z7xSQ3SXLQU}.

\bibitem[Shumailov et~al.(2023)Shumailov, Shumaylov, Zhao, Gal, Papernot, and
  Anderson]{shumailov2023curse}
I.~Shumailov, Z.~Shumaylov, Y.~Zhao, Y.~Gal, N.~Papernot, and R.~Anderson.
\newblock The curse of recursion: Training on generated data makes models
  forget, 2023.

\bibitem[Stawinski()]{torchbreak2stawinski}
J.~Stawinski.
\newblock Playing with fire – how we executed a critical supply chain attack
  on pytorch.
\newblock URL
  \url{https://johnstawinski.com/2024/01/11/playing-with-fire-how-we-executed-a-critical-supply-chain-attack-on-pytorch/}.
\newblock Accessed: 2024-01-11.

\bibitem[Szegedy et~al.(2014)Szegedy, Liu, Jia, Sermanet, Reed, Anguelov,
  Erhan, Vanhoucke, and Rabinovich]{szegedy2014going}
C.~Szegedy, W.~Liu, Y.~Jia, P.~Sermanet, S.~Reed, D.~Anguelov, D.~Erhan,
  V.~Vanhoucke, and A.~Rabinovich.
\newblock Going deeper with convolutions, 2014.

\bibitem[Tan and Le(2021)]{tan2021efficientnetv2}
M.~Tan and Q.~Le.
\newblock Efficientnetv2: Smaller models and faster training.
\newblock In M.~Meila and T.~Zhang, editors, \emph{Proceedings of the 38th
  International Conference on Machine Learning}, volume 139 of
  \emph{Proceedings of Machine Learning Research}, pages 10096--10106. Pmlr,
  18--24 Jul 2021.
\newblock URL \url{https://proceedings.mlr.press/v139/tan21a.html}.

\bibitem[TheHackerNews()]{samsung}
TheHackerNews.
\newblock Samsung printer having secret admin account backdoor.
\newblock URL
  \url{https://thehackernews.com/2012/11/samsung-printer-having-secret-admin.html}.
\newblock Accessed: 2023-12-18.

\bibitem[Thompson(1984)]{thompson1984reflections}
K.~Thompson.
\newblock Reflections on trusting trust.
\newblock \emph{Commun. ACM}, 27\penalty0 (8):\penalty0 761–763, aug 1984.
\newblock ISSN 0001-0782.
\newblock \doi{10.1145/358198.358210}.
\newblock URL \url{https://doi.org/10.1145/358198.358210}.

\bibitem[TrailOfBits(2021)]{fickling}
TrailOfBits.
\newblock Fickling, 2021.
\newblock URL \url{https://github.com/trailofbits/fickling}.

\bibitem[TrailOfBits(2023)]{semgrep}
TrailOfBits.
\newblock Semgrep, 2023.
\newblock URL \url{https://github.com/trailofbits/semgrep-rules}.

\bibitem[Travers(2021)]{lobotoml}
A.~Travers.
\newblock Lobotoml, 2021.
\newblock URL \url{https://github.com/alkaet/LobotoMl/}.

\bibitem[Wang et~al.(2019)Wang, Yao, Shan, Li, Viswanath, Zheng, and
  Zhao]{wang2019neuralcleans}
B.~Wang, Y.~Yao, S.~Shan, H.~Li, B.~Viswanath, H.~Zheng, and B.~Y. Zhao.
\newblock Neural cleanse: Identifying and mitigating backdoor attacks in neural
  networks.
\newblock In \emph{2019 IEEE Symposium on Security and Privacy (SP)}, pages
  707--723, 2019.
\newblock \doi{10.1109/sp.2019.00031}.

\bibitem[Warnecke et~al.(2023)Warnecke, Speith, Möller, Rieck, and
  Paar]{warnecke2023evil}
A.~Warnecke, J.~Speith, J.-N. Möller, K.~Rieck, and C.~Paar.
\newblock Evil from within: Machine learning backdoors through hardware
  trojans, 2023.

\bibitem[Xia et~al.(2023)Xia, Bi, Xing, Lu, and Zhu]{xia2023empirical}
B.~Xia, T.~Bi, Z.~Xing, Q.~Lu, and L.~Zhu.
\newblock An empirical study on software bill of materials: Where we stand and
  the road ahead.
\newblock \emph{arXiv preprint arXiv:2301.05362}, 2023.

\bibitem[Young()]{torchbreak3young}
M.~Young.
\newblock Zuckerpunch - abusing self hosted github runners at facebook.
\newblock URL \url{https://marcyoung.us/post/zuckerpunch/}.
\newblock Accessed: 2023-12-18.

\bibitem[Zhao et~al.(2022)Zhao, Gao, Shumailov, Fusi, and
  Mullins]{zhao2022rapid}
Y.~Zhao, X.~Gao, I.~Shumailov, N.~Fusi, and R.~D. Mullins.
\newblock Rapid model architecture adaption for meta-learning.
\newblock In A.~H. Oh, A.~Agarwal, D.~Belgrave, and K.~Cho, editors,
  \emph{Advances in Neural Information Processing Systems}, 2022.
\newblock URL \url{https://openreview.net/forum?id=Yq6g9xluV0}.

\bibitem[Zhou et~al.(2023)Zhou, Gao, Wu, Grundy, Chen, Chen, and
  Li]{zhou2023model}
M.~Zhou, X.~Gao, J.~Wu, J.~C. Grundy, X.~Chen, C.~Chen, and L.~Li.
\newblock Model obfuscation for securing deployed neural networks, 2023.
\newblock URL \url{https://openreview.net/forum?id=ib482K6HQod}.

\end{thebibliography}
